\documentclass[sigplan,twocolumn,screen]{acmart}

\renewcommand\footnotetextcopyrightpermission[1]{}
\usepackage{colortbl}
\usepackage[dvipsnames,table,xcdraw]{xcolor}
\usepackage[most]{tcolorbox}
\usepackage{booktabs}
\usepackage{tabularx}
\usepackage{multirow}
\usepackage{algorithm}
\usepackage{ulem}
\usepackage{algpseudocode}
\usepackage{soul}
\usepackage{balance} 
\usepackage{tikz}
\usepackage{enumitem}
\usepackage{graphicx}
\usepackage{flushend}
\usepackage{helvet}
\usepackage{balance}
\newcommand*\circled[1]{\tikz[baseline=(char.base)]{
            \node[shape=circle,fill,inner sep=1pt] (char) {\footnotesize \textcolor{white}{#1}};}}

\usepackage{hyperref}
\hypersetup{
 colorlinks=true,      
 linkcolor=blue,       
 citecolor=red,    
 filecolor=cyan,       
 urlcolor=magenta          
}

\newcounter{insight}

\newtcolorbox{insightbox}{
  enhanced,
  breakable,
  colback=gray!5,
  colframe=black!55,
  boxrule=0.4pt,
  arc=1pt,
  left=4pt,
  right=4pt,
  top=2pt,
  bottom=2pt,
  before skip=4pt,
  after skip=4pt,
  before upper={
    \refstepcounter{insight}%
    \textbf{Insight~\theinsight.}\enspace
  }
}

\newcommand{\lastlinemaybreak}{{\widowpenalty=0\par}}
\newcommand{\firstlinemaybreak}{{\clubpenalty=0\par}}

\begin{document}

\title{RankServe: Rank-Aware CPU–GPU Scheduling for Memory-Constrained MoE Serving}

\author{Hengyi Zhu}
\email{hzhu37@stevens.edu}
\affiliation{%
  \institution{Stevens Institute of Technology}
  \city{Hoboken}
  \state{New Jersey}
  \country{USA}
}

\author{Zhendong Mi}
\email{zmi2@stevens.edu}
\affiliation{%
  \institution{Stevens Institute of Technology}
  \city{Hoboken}
  \state{New Jersey}
  \country{USA}
}

\author{Pu Zhao}
\email{p.zhao@northeastern.edu}
\affiliation{%
  \institution{Northeastern University}
  \city{Boston}
  \state{Massachusetts}
  \country{USA}
}

\author{Yanzhi Wang}
\email{yanz.wang@northeastern.edu}
\affiliation{%
  \institution{Northeastern University}
  \city{Boston}
  \state{Massachusetts}
  \country{USA}
}

\author{Hao Wang}
\email{hwang9@stevens.edu}
\affiliation{%
  \institution{Stevens Institute of Technology}
  \city{Hoboken}
  \state{New Jersey}
  \country{USA}
}

\author{Shaoyi Huang}
\correspondingauthor
\email{shuang59@stevens.edu}
\affiliation{%
  \institution{Stevens Institute of Technology}
  \city{Hoboken}
  \state{New Jersey}
  \country{USA}
}

\begin{abstract}
Mixture-of-Experts (MoE) models scale model capacity through sparse expert activation, but their large expert pools impose substantial memory requirements during inference. Low-rank compression reduces the memory footprint of Mixture-of-Experts (MoE) models. However, the heterogeneous expert ranks create unequal storage, transfer, and execution costs. Under limited GPU memory, expert popularity and transfer cost alone cannot determine which experts are most valuable to retain. We present RankServe, a CPU--GPU serving system for compressed MoE checkpoints with fixed expert ranks. Its key insight is to value GPU residency against the cheaper nonresident path, CPU execution or weight transfer followed by GPU execution, under each expert's rank and token load. RankServe aggregates the resulting estimated savings over routing demand and prioritizes experts by savings per byte. It updates residency from recent routing history and selects miss paths using current token loads, allowing temporary GPU execution without cache admission. A variable-rank runtime executes heterogeneous factors through grouped matrix multiplications without common-rank padding. Implemented on vLLM, RankServe is evaluated on three MoE models against six serving systems on a single CPU--GPU platform. Across five batch sizes from 1 to 32, it achieves per-model geometric-mean speedups of $1.69$--$2.29\times$ in time to first token (TTFT) and $1.71$--$2.48\times$ in time per output token (TPOT), relative to the fastest baseline for each configuration and metric.
\end{abstract}

\maketitle
\pagestyle{plain}

\section{Introduction}
\label{sec:introduction}


Mixture-of-Experts (MoE) architectures expand model capacity by activating only a subset of experts for each token, avoiding a proportional increase in per-token computation~\cite{fedus2022switch,jiang2024mixtral}. This computational sparsity, however, does not eliminate the need to store the complete expert pool.
When expert weights exceed GPU memory, offloading to CPU memory requires deciding which weights to retain on the GPU, which weights to transfer, and where to execute each activated expert~\cite{kamahori2025fiddler}.
Low-rank compression reduces expert storage by replacing dense weight matrices with smaller factors that can be stored, transferred, and executed directly~\cite{wang2025svd,li2025structured}. Because experts differ in their sensitivity to compression, recent methods assign non-uniform ranks under a parameter budget to preserve model quality~\cite{mi2026effective}. The resulting experts have different weight footprints and factor dimensions, even within the same MoE layer. Such a compressed checkpoint can still exceed GPU memory, leaving the serving system to manage both heterogeneous expert
costs and changing routing demand.


\begin{figure}[!t]
\centering
\begin{tikzpicture}[
    x=\columnwidth,
    y=1cm,
    every node/.style={
        align=center,
        inner sep=1pt,
        outer sep=0pt,
        font=\fontfamily{ptm}\fontsize{8}{9.3}\selectfont
    },
    head/.style={
        anchor=north,
        font=\fontfamily{ptm}\fontsize{8.5}{9.7}\selectfont\bfseries
    },
    papers/.style={
        anchor=north,
        font=\fontfamily{ptm}\fontsize{8}{9.3}\selectfont
    },
    branch/.style={
        draw=blue!55!cyan,
        line width=.4pt
    },
    focus/.style={
        draw=red!75!black,
        line width=.75pt
    }
]

\node[
    anchor=north,
    font=\fontfamily{ptm}\fontsize{9.5}{11}\selectfont\bfseries
] at (.50,0)
    {Memory-Constrained MoE Serving};

\draw[branch]
    (.50,-.43)--(.50,-.58)
    --(.14,-.58)--(.14,-.74);
\draw[branch]
    (.455,-.58)--(.455,-.74);
\draw[focus]
    (.50,-.58)--(.825,-.58)--(.825,-.74);

\node[head,text width=.27\columnwidth] at (.14,-.82)
    {Computation\\Adaptation};
\node[head,text width=.33\columnwidth] at (.455,-.82)
    {Movement and\\Execution Scheduling};
\node[head,text width=.32\columnwidth] at (.825,-.82)
    {Memory\\Allocation};

\node[papers,text width=.27\columnwidth] at (.14,-1.60) {
    AdapMoE\cite{zhong2024adapmoe}\\
    HOBBIT\cite{tang2024hobbit}\\
    SPICE\cite{lyu2026spice}\\
    MoE-APEX\cite{tang2026moe}
};

\node[papers,text width=.33\columnwidth] at (.455,-1.60) {
    Mixtral-Offload\cite{eliseev2023fast}\\
    MoE-Infinity\cite{xue2024moe}\\
    ProMoE\cite{song2024promoe}\\
    FineMoE\cite{yu2026taming}\\
    KTransformers\cite{chen2025ktransformers}\\
    MoE-Lightning\cite{cao2025moe}
};

\draw[branch]
    (.825,-1.57)--(.825,-1.76)
    --(.72,-1.76)--(.72,-1.94);
\draw[focus]
    (.825,-1.76)--(.91,-1.76)--(.91,-1.94);

\node[head,text width=.16\columnwidth] at (.72,-2.02)
    {Pool\\Budgets};
\node[head,text width=.18\columnwidth] at (.91,-2.02)
    {Expert\\Residency};

\node[papers,text width=.16\columnwidth] at (.72,-2.80)
    {WiSP\cite{zhang2026wisp}};

\node[papers,text width=.18\columnwidth] at (.91,-2.80) {
    MoE-SpAc\\
    \cite{li2026moe}\\[1pt]
    FreeToken\\
    \cite{yang2026freetoken}
};

\draw[branch]
    (.91,-4.18)--(.91,-4.38)
    --(.17,-4.38)--(.17,-4.56);
\draw[branch]
    (.50,-4.38)--(.50,-4.56);
\draw[focus]
    (.91,-4.18)--(.91,-4.38)
    --(.83,-4.38)--(.83,-4.56);

\node[head,text width=.30\columnwidth] at (.17,-4.66)
    {Demand-Based\\Priority};
\node[head,text width=.30\columnwidth] at (.50,-4.66)
    {Precision-Aware\\Miss Costs};
\node[
    head,
    text=red!75!black,
    text width=.32\columnwidth
] at (.83,-4.66)
    {Rank-Dependent\\Execution Savings};

\node[papers,text width=.30\columnwidth] at (.17,-5.46) {
    Fiddler\cite{kamahori2025fiddler}\\
    HybriMoE\cite{zhong2025hybrimoe}
};
\node[papers,text width=.30\columnwidth] at (.50,-5.46) {
    HOBBIT\cite{tang2024hobbit}\\
    MoE-APEX\cite{tang2026moe}
};

\node[
    anchor=center,
    text=red!75!black,
    font=\fontfamily{ptm}\fontsize{9.5}{11}\selectfont\bfseries
] at (.83,-5.78)
    {RankServe};

\end{tikzpicture}
\caption{Design space of memory-constrained MoE serving.}
\vspace{-0.2in}
\label{fig:design-space}
\end{figure}
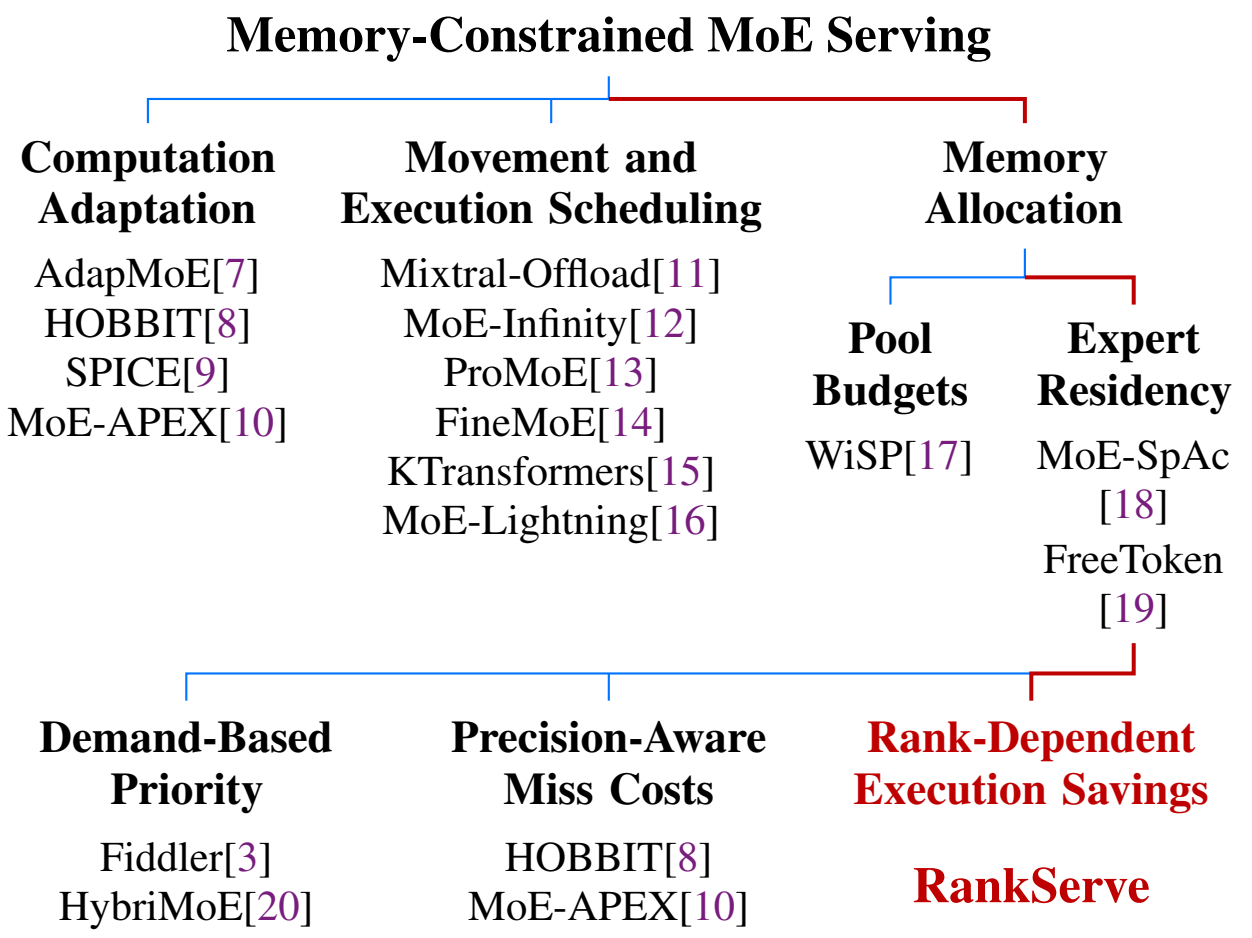

Figure~\ref{fig:design-space} demonstrates existing memory-constrained MoE serving by three decision dimensions: \textit{computation adaptation}, \textit{movement and execution scheduling}, and \textit{memory allocation}.
\textit{Computation adaptation} adjusts expert selection or precision, or substitutes low-rank surrogates for selected expert misses~\cite{zhong2024adapmoe,tang2024hobbit,yang2026faster,liu2025bandwidth}.
\textit{Movement and execution scheduling} determines when weights move and where and when expert computations run, using on-demand loading~\cite{eliseev2023fast,xue2024moe}, predictive prefetching~\cite{song2024promoe,yu2026taming}, and CPU--GPU placement and overlap~\cite{cao2025moe,chen2025ktransformers,kamahori2025fiddler,zhong2025hybrimoe}.
\textit{Memory allocation} determines how GPU capacity is divided among memory pools and which expert weights remain resident.
%
It spans several levels: WiSP~\cite{zhang2026wisp} allocates capacity between expert and KV-cache pools, while HOBBIT~\cite{tang2024hobbit} and
MoE-APEX~\cite{tang2026moe} incorporate precision-related costs into expert-cache management. MoE-SpAc~\cite{li2026moe} and FreeToken~\cite{yang2026freetoken} also coordinate expert caching with heterogeneous execution. 
Within this design space, we focus on expert residency for a fixed checkpoint with heterogeneous expert ranks. The question is how to translate each expert's demand, memory footprint, and execution alternatives into a GPU-residency priority.


Our key insight is that \textit{an expert's value in GPU memory depends on how cheaply it can execute without GPU residency}. Rank determines weight size and transfer volume, while rank and routed-token load jointly affect CPU and GPU execution times. These costs need not scale together: smaller factors reduce transferred bytes, but GPU execution at small token loads can remain dominated by fixed overheads. Consequently, frequently activated experts can still offer little benefit from residency when CPU execution is inexpensive. Demand alone cannot determine residency value, and a transfer-only miss model also overestimates that value whenever CPU execution is cheaper than fetching the expert
for GPU execution. Residency should therefore be evaluated against the cheaper nonresident path and normalized by the bytes required to retain the expert.

We present RankServe, a serving system that allocates GPU residency by \textit{rank-dependent execution savings} for heterogeneous low-rank MoE experts under a fixed GPU memory budget. RankServe takes a compressed checkpoint as input and preserves its assigned ranks and router-selected expert computations. For each expert access, it estimates the time saved by resident GPU execution relative to the cheaper of direct CPU execution and weight transfer followed by GPU execution. It aggregates these estimated savings over routing demand and prioritizes experts by savings per byte, accounting for both how often an expert is activated and how many tokens it receives at each access. This places RankServe at the expert-residency level of memory allocation, where the value assigned to GPU memory explicitly depends on the available miss-execution paths.

RankServe initializes residency priorities from calibration traces and updates them using recent routing history. For each cache miss, its online scheduler selects an execution path using the current
routed-token load and profiled costs. Temporarily fetched weights use separate storage, allowing a GPU-executed expert to remain outside the resident cache. To execute differently shaped factors, a
variable-rank GPU runtime groups expert computations without padding them to a common rank. Our contributions are as follows:%
\looseness=1 

\begin{itemize}
\setlength{\itemsep}{6pt}
    \setlength{\parsep}{0pt}
    \setlength{\parskip}{0pt}
    \item 
    We formulate residency value as cumulative estimated execution savings per byte relative to the cheaper CPU or fetch-and-GPU path, accounting for rank-dependent costs, routed-token loads, and unequal expert footprints.
    \item 
    We coordinate residency updates from recent routing history with miss-path selection from current token loads. Temporary GPU execution is separated from cache admission, preserving cache capacity for experts with higher residency value.
    \item 
    We develop a GPU runtime that groups heterogeneous low-rank projections, handling rank-dependent output widths and reduction lengths with one grouped launch per stage and avoiding common-rank padding.
    \item 
    We implement RankServe on vLLM~\cite{kwon2023efficient} and evaluate three MoE models against six serving systems across batch sizes and GPU weight budgets. Across the five batch sizes in the main latency experiment, per-model geometric-mean speedups are $1.69$--$2.29\times$ for TTFT and $1.71$--$2.48\times$ for TPOT, relative to the fastest baseline for each configuration and metric.
\end{itemize}

\section{Preliminaries}
\label{sec:preliminaries}

\subsection{MoE Architecture}
\label{sec:bg-moe}

Most modern LLMs use decoder-only Transformers for autoregressive generation~\cite{touvron2023llamaopenefficientfoundation}. Prefill processes prompt tokens in parallel, while subsequent standard decoding generates one token per sequence per iteration using the same Transformer layers~\cite{kwon2023efficient}. MoE-based LLMs replace some or all feed-forward network (FFN) layers with sparsely activated experts~\cite{fedus2022switch,jiang2024mixtral}.
With top-$K$ routing, each MoE layer's router selects $K$ experts for token representation $h_t$, and the layer combines their outputs as
\begin{equation}
  y_t = \sum_{i\in\mathcal{T}_t} p_{t,i}E_i(h_t),
  \label{eq:moe-routing}
\end{equation}
where $\mathcal{T}_t$ is the selected expert set, $E_i$ denotes expert $i$, and $p_{t,i}$ is its routing weight. Sparse activation reduces per-token expert computation but does not reduce the storage required for the full expert pool. At each layer, routing determines expert selection and token loads; the inference system must make selected experts available and schedule their execution.

\subsection{Low-Rank MoE Compression and Heterogeneous Expert Rank Allocation}
\label{sec:bg-svd}

\begin{figure*}[!ht]
    \centering
    \includegraphics[width=0.95\textwidth]{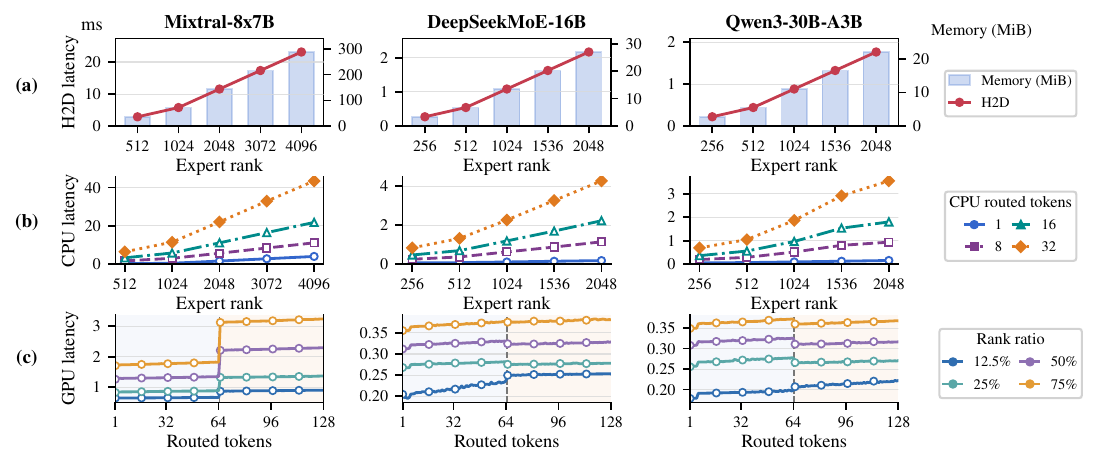}
    \vspace{-0.2in}
    \caption{
    (a) Expert-weight footprint and H2D transfer latency at different retained ranks; (b) CPU execution latency versus rank at different routed-token loads; (c) GPU execution latency versus routed-token load at different rank ratios. CPU and GPU execution exhibit different scaling behavior across the evaluated configurations.
    }
    \label{fig:obs1}
    \vspace{-0.15in}
\end{figure*}

Low-rank compression approximates a weight matrix
$W\in\mathbb{R}^{d_{\mathrm{out}}\times d_{\mathrm{in}}}$ using truncated SVD:
$W \approx U_r\Sigma_rV_r^{\mathsf T}=AB,$
where $A=U_r\Sigma_r\in\mathbb{R}^{d_{\mathrm{out}}\times r}$ and
$B=V_r^{\mathsf T}\in\mathbb{R}^{r\times d_{\mathrm{in}}}$.
This reduces the parameter count from $d_{\mathrm{out}}d_{\mathrm{in}}$ to
$r(d_{\mathrm{out}}+d_{\mathrm{in}})$, with inference computed as
$Wx\approx A(Bx)$~\cite{wang2025svd}.
For MoEs, expert matrices can be factorized independently, with further savings from sharing factors~\cite{li2025structured,liu2025molae} or bases~\cite{chen2026mobe} across experts.

Since weight matrices differ in compression sensitivity, different ranks can be assigned to different experts.
Prior methods allocate heterogeneous ranks using truncation loss~\cite{wang2025svd2}, calibration importance~\cite{yang2024moe}, or activation and weight whitening metrics~\cite{huang2025milo}.
We adopt RFID-MoE's allocation~\cite{mi2026effective}, which groups experts by routing frequency and determines each group's rank using both routing frequency and effective rank.
Let $F_g$ and $E_g$ denote the normalized routing frequency and effective rank of group $g$. The group importance and retained rank are
\begin{equation}
C_g=\xi E_g+(1-\xi)F_g,\quad
r_g=\max\left(1,\left\lfloor
R_{\mathrm{total}}\frac{C_g}{\sum_{g'}C_{g'}}
\right\rfloor\right),
\label{eq:rank-allocation}
\end{equation}
where $\xi$ balances the two metrics and $R_{\mathrm{total}}$ is the rank budget determined by the target compression ratio. All experts within a group share the assigned rank.
Consequently, an expert $i$ has factors of dimensions
$d_{\mathrm{out}}\times r_i$ and $r_i\times d_{\mathrm{in}}$.
Experts with identical original dimensions can therefore have different compressed sizes and computation shapes, resulting in heterogeneous storage, transfer, and execution costs.
RankServe takes these offline-assigned ranks as fixed inputs and manages the resulting variable-rank experts during inference.

\section{Limitations, Observations and Opportunities}
\label{sec:observations}

\begin{figure*}[!t]
    \centering
    \includegraphics[width=0.99\textwidth]{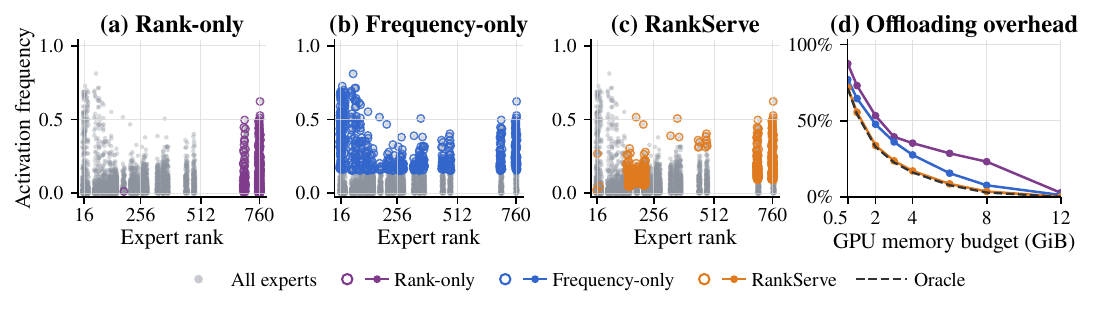}
    \vspace{-0.3in}
    \caption{
    GPU placement for single-request Qwen3-30B-A3B decoding. In the first three subfigures, gray dots represent all experts, and colored dots indicate experts selected for GPU residency under a 2 GiB GPU cache budget. The final subfigure compares the remaining normalized offloading overhead across GPU cache budgets; lower is better. 
    }
    \vspace{-0.15in}
    \label{fig:obs2-placement}
\end{figure*}

Heterogeneous rank allocation can better preserve model quality under a fixed parameter budget by considering differences among experts~\cite{mi2026effective}. 
We take the resulting compressed checkpoint and its assigned ranks as fixed inputs and study how to serve it efficiently under limited GPU memory.
Profiling Mixtral-8 $\times$7B, DeepSeekMoE-16B, and Qwen3-30B-A3B reveals two considerations for expert management: the latency savings from GPU residency relative to its memory cost, and how routing demand changes both these savings and the preferred execution path.

\vspace{-0.2in}
\subsection{Limitations of Existing Memory-Constrained MoE Serving Solutions}

Figure~\ref{fig:design-space} organizes existing solutions along three overlapping decision dimensions: computation adaptation, movement and execution scheduling, and memory allocation. 

\noindent\textbf{Computation adaptation.}
Methods such as AdapMoE, HOBBIT, MoE-APEX, and SPICE reduce resource costs by adjusting expert selection, precision, or computation~\cite{zhong2024adapmoe,tang2024hobbit,tang2026moe,lyu2026spice}. These adaptations reduce resource requirements, but do not eliminate differences in expert memory footprints and execution costs. For a compressed checkpoint with fixed heterogeneous ranks, a frequently activated expert may still be inexpensive to execute on the CPU, whereas another expert may benefit more from GPU residency but occupy more memory. The serving system must therefore determine how much execution time retaining each expert saves per byte of GPU memory, given its rank and token load.

\noindent\textbf{Movement and execution scheduling.}
Expert-offloading systems reduce loading stalls through caching and prefetching~\cite{eliseev2023fast,xue2024moe,song2024promoe,yu2026taming}. Fiddler and HybriMoE use workload information to choose between CPU execution and weight transfer followed by GPU execution~\cite{kamahori2025fiddler,zhong2025hybrimoe}. These mechanisms can make cache misses cheaper, which also reduces the benefit of keeping an expert resident. For example, an expert whose weights are expensive to transfer may still execute cheaply on the CPU when it receives few tokens. A residency priority based only on its transfer cost would then overestimate the time saved by caching it. Residency decisions must therefore account for the cheaper nonresident execution path under the expert's rank and token load, and compare the resulting savings per byte of GPU memory.

\noindent\textbf{Memory allocation.}
WiSP determines how GPU memory is divided between expert weights and the KV cache~\cite{zhang2026wisp}. Within the expert cache, the system must still decide which experts to retain. Fiddler uses offline-profiled expert popularity, while FreeToken uses recency-based replacement~\cite{kamahori2025fiddler,yang2026freetoken}. MoE-SpAc combines speculative activation utilities with CPU--GPU workload balancing, but its formulation uses a common per-expert memory size and unit execution and I/O costs~\cite{li2026moe}. HOBBIT and MoE-APEX further account for precision-related costs in cache management~\cite{tang2024hobbit,tang2026moe}.

With heterogeneous ranks, however, experts occupy different amounts of memory, and a cache hit can save different amounts of execution time depending on the expert's rank and token load. A frequently activated expert may be cheap to execute on the CPU, so retaining it yields little benefit despite its frequent reuse. A larger expert may save more time per access, yet consume space that could retain several smaller experts with greater combined savings. Allocating the expert cache therefore requires comparing cumulative execution savings against memory consumption, rather than treating activation frequency or per-expert costs as sufficient measures of residency value.

\subsection{Observations and Opportunities}

\noindent\textbf{Observation 1: Demand-based priorities can misallocate GPU memory across heterogeneous-rank experts.}
A frequently activated expert can be inexpensive to serve without residency, while a higher-rank expert consumes more cache space even when caching avoids greater transfer or CPU execution costs. Neither activation frequency nor rank alone establishes which experts use GPU memory best.

Figure~\ref{fig:obs1} profiles how rank and routed-token load affect serving costs across three models. Expert-weight footprints and host-to-device (H2D) transfer latencies increase with rank (Figure~\ref{fig:obs1}(a)). CPU execution latency also increases with rank, more steeply at higher routed-token loads (Figure~\ref{fig:obs1}(b)). GPU execution latency depends on both rank and token load, with model-dependent scaling (Figure~\ref{fig:obs1}(c)). These profiles show why residency decisions must account for both an expert's serving cost and the memory needed to retain it.



Figure~\ref{fig:obs2-placement} compares residency policies for single-request Qwen3-30B-A3B decoding. Under the same 2\,GiB GPU cache budget, rank-only and frequency-only favor higher ranks and activation frequencies, respectively (Figures~\ref{fig:obs2-placement}(a)--(c)). Using profiled latencies, demand, and weight footprints, RankServe achieves lower normalized offloading overhead than both(Figures~\ref{fig:obs2-placement}(d)). These simplified baselines isolate priority limitations; they do not represent existing systems' full caching and scheduling mechanisms.\firstlinemaybreak 


\noindent\textbf{Opportunity 1: Allocate GPU cache capacity by estimated execution savings per byte.}
For each access, estimate the time saved by resident GPU execution relative to the cheaper of CPU execution and weight transfer followed by GPU execution. Sum these savings over the workload window and divide by the expert's weight footprint to obtain a residency priority. 

\begin{figure*}[!t]
    \centering
    \includegraphics[width=1\textwidth]{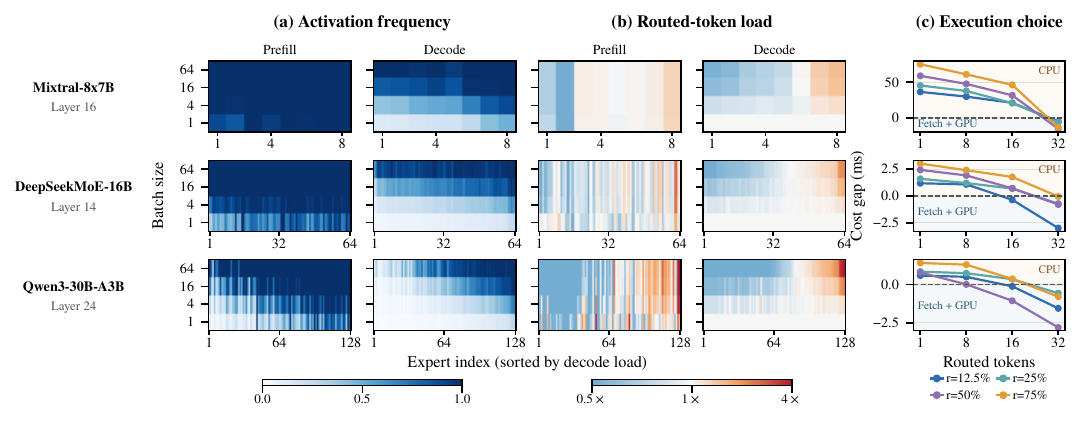}
    \vspace{-0.4in}
    \caption{
    (a) Expert activation frequency during prefill and decode across batch sizes. (b) Relative routed-token load, conditioned on each expert being active, with experts ordered as in (a). (c) Fetch-and-GPU latency minus CPU-path latency across expert ranks and routed-token loads; positive values favor the CPU path, and negative values favor fetch-and-GPU execution.
    }
    \vspace{-0.1in}
    \label{fig:obs2}
\end{figure*}

\noindent\textbf{Observation 2: Token load changes both the cheaper miss path and the value of residency.}
Expert demand has two distinct dimensions: activation frequency is the fraction of iterations in which an expert receives at least one token, whereas routed-token load is the number of tokens assigned to it in an iteration. Frequency measures how often an expert is needed; token load determines the amount of work at each access. Figure~\ref{fig:obs2}(a) shows that activation frequencies differ between prefill and decode and generally increase with batch size. Routed-token loads remain uneven across experts, even when conditioned on activation (Figure~\ref{fig:obs2}(b)). Frequent activation therefore does not imply a large load per access, so frequency alone cannot identify the cheaper nonresident execution path.
Figure~\ref{fig:obs2}(c) plots fetch-and-GPU latency minus CPU-path latency. The CPU path retains weights in host memory but incurs activation transfers and synchronization; fetch-and-GPU transfers weights before execution. In the evaluated configurations, CPU execution is faster at small loads, while fetch-and-GPU becomes preferable at larger loads, with model- and rank-dependent crossovers.


When the CPU path is cheaper, the estimated residency benefit is CPU-path latency minus resident GPU latency. When fetch-and-GPU execution is cheaper, the benefit equals the avoided weight-transfer time. Under this additive cost model, the per-access saving is bounded above by the weight-transfer time, even if CPU latency continues to grow with token load. A transfer-only estimate therefore overvalues accesses served more cheaply on the CPU, while a frequency-only priority treats accesses with different savings alike. Frequency determines how often savings accumulate, and the routed-token load determines their value through the cheaper nonresident path.

\noindent\textbf{Opportunity 2: Use the same cost model for residency updates and miss dispatch.}
Use recent routed-token histories to update savings-per-byte residency priorities, and current token loads to choose between feasible CPU and fetch-and-GPU paths for cache misses. An expert fetched for the current iteration can use temporary GPU storage without entering the resident cache. This separates the immediate execution choice from the decision to spend cache capacity on future reuse. Realizing these decisions also requires executing the resulting GPU workload with its assigned low-rank factor shapes, motivating the variable-rank runtime described later.

\section{RankServe Overview}
\label{sec:overview}

\subsection{Objectives and Challenges}
\label{sec:objectives}

\begin{figure}[!t]
    \centering
    \includegraphics[width=0.96\columnwidth]{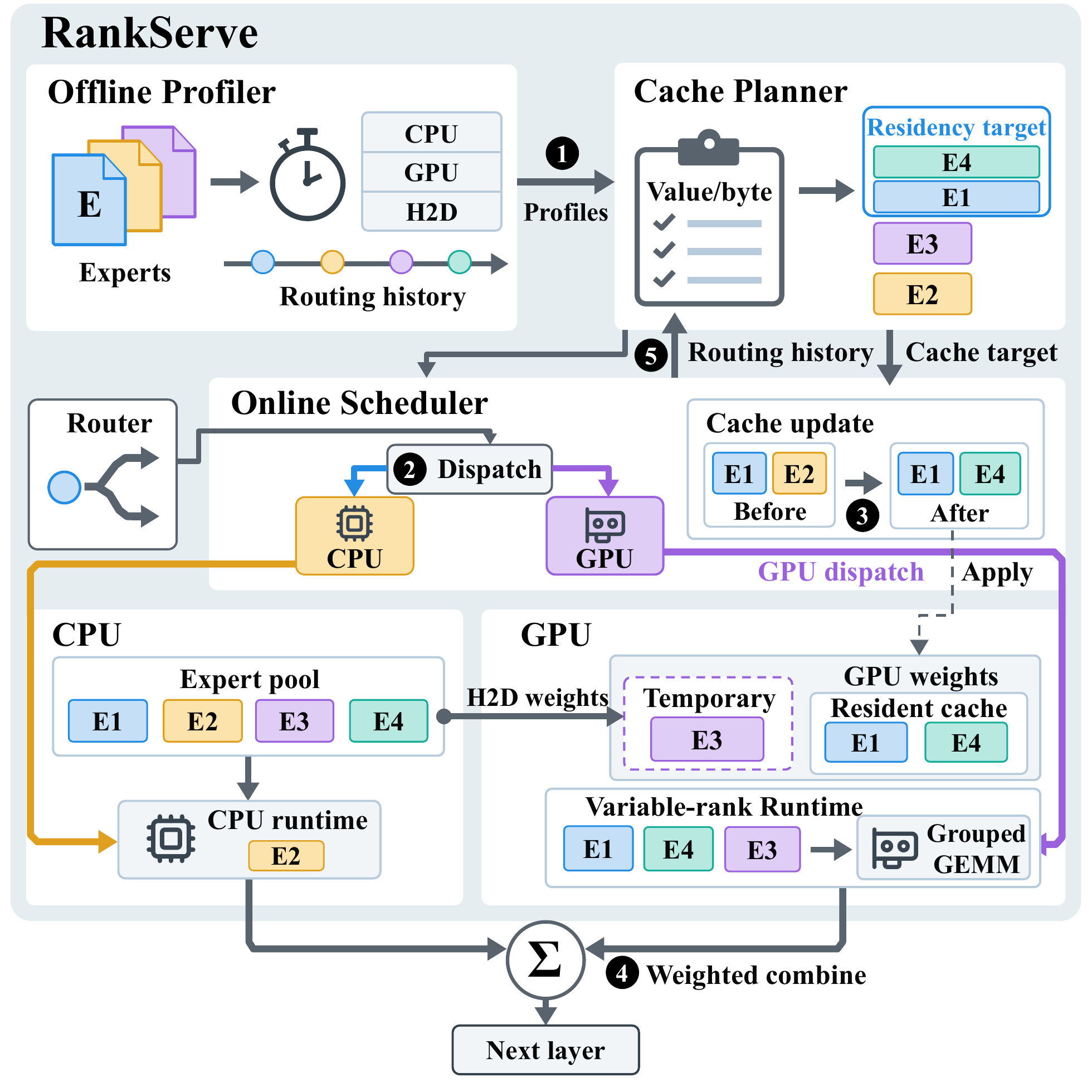}
    \vspace{-0.1in}
    \caption{The overall workflow of RankServe. 
    }
    \vspace{-0.2in}
    \label{fig:framework}
\end{figure}

RankServe targets low-rank-compressed MoE models. Given a fixed checkpoint, its \textbf{objective} is to reduce inference latency through coordinated GPU residency and CPU--GPU execution under a fixed expert-cache budget, while preserving assigned ranks and router selections. Adapting to changing demand and efficiently executing heterogeneous factor shapes introduce the following design challenges.

\textbf{Challenge 1: How can a fixed cache budget yield the greatest latency savings?}
Heterogeneous ranks produce unequal memory footprints, transfer costs, and execution costs. Frequent reuse may yield little benefit when CPU execution is cheap, while greater savings per access may require more memory. Allocation must compare cumulative savings against unequal memory costs, with each access valued relative to its cheaper nonresident execution path.


\textbf{Challenge 2: How can residency and execution sustain low latency under changing demand?}
Residency depends on reuse across iterations, whereas miss dispatch depends on the current token load. Demand changes can switch the preferred miss path and reorder residency priorities. These decisions must be coordinated across timescales: fetching an expert for immediate GPU execution does not necessarily justify retaining it for future accesses.

\subsection{Architecture and Workflow}
\label{sec:architecture}

Figure~\ref{fig:framework} illustrates the architecture and workflow of Rank-Serve, which consists of four main components:

\noindent\textbf{Offline Profiler.}
The profiler characterizes the demand and serving costs of compressed experts. It records expert activation frequencies and routed-token loads from calibration inputs, and measures CPU execution-path latency, GPU execution latency, and host-to-device (H2D) weight-transfer latency on the target hardware. Together with expert weight sizes, these profiles provide the basis for cache planning and execution scheduling.

\noindent\textbf{Cache Planner.}
The planner determines which experts should reside in GPU memory under the cache budget. It combines expert demand with profiled costs to estimate the latency savings per byte of GPU residency and produces a cache target. Online routing and cost information informs updates to this target as expert demand changes.

\noindent\textbf{Online Scheduler.}
The scheduler receives routing assignments and coordinates expert dispatch and cache updates. It dispatches resident experts to the GPU and selects between CPU execution and fetch-and-GPU execution for nonresident experts. It also applies cache admissions and evictions according to the planner's target.

\noindent\textbf{Variable-Rank GPU Runtime.}
The runtime executes GPU-dispatched experts using their compressed factors. It supports different retained ranks within grouped GEMMs, using per-expert shape and weight-offset metadata to avoid padding all experts to a common rank. Both cached experts and temporarily fetched experts use this runtime.

RankServe follows the workflow below to coordinate expert residency and execution:

\noindent\textbf{Step \circled{1}: Offline profiling and initial cache planning.}
Before inference, the profiler collects expert demand and cost measurements for the compressed checkpoint. The cache planner uses these measurements to rank experts by expected latency savings per byte and establish the initial GPU residency target. The complete expert pool remains available in CPU memory.

\noindent\textbf{Step \circled{2}: Routing and execution-path selection.}
At each MoE layer, the router identifies the activated experts and their routed-token loads. The scheduler checks their GPU residency and selects an execution path for each expert. Resident experts execute on the GPU, while nonresident experts either execute on the CPU or have their weights transferred for GPU execution. These decisions account for the current workload, expert weight sizes, and profiled execution and transfer costs.

\noindent\textbf{Step \circled{3}: Cache update and weight preparation.}
The scheduler applies the cache target within the GPU memory budget, retaining selected experts and evicting lower-priority weights as needed. Weights fetched for GPU execution can also use temporary storage without entering the resident cache. In Figure~\ref{fig:framework}, the cache update replaces E2 with E4 while retaining E1; E3 is loaded into temporary GPU storage, and E2 remains available for CPU execution.

\noindent\textbf{Step \circled{4}: CPU--GPU execution and output combination.}
The CPU runtime executes E2 using host-resident weights, while the variable-rank GPU runtime executes E1, E4, and E3 through grouped GEMMs. CPU execution transfers the required activations between devices, whereas fetch-and-GPU execution transfers expert weights. Independent transfers and computation can overlap once their input dependencies are satisfied. The expert outputs are combined using the routing weights to produce the MoE layer output.

\begin{table}[!t]
    \centering
    \caption{Notation used in residency planning and online scheduling.}
    \vspace{-0.1in}
    \label{tab:notation}
    \small
    \setlength{\tabcolsep}{4pt}
    \renewcommand{\arraystretch}{1.08}
    \begin{tabular}{@{}lp{0.82\columnwidth}@{}}
        \toprule
        Symbol & Definition \\
        \midrule
        $i=(\ell,j)$ & Expert $j$ in MoE layer $\ell$ \\
        $\mathcal{E},\,\mathcal{E}_\ell$ & Expert sets across all layers and in layer $\ell$ \\
        $\mathcal{W},\,k$ & Workload window and iteration index \\
        $\mathcal{W}_{\mathrm{profile}}$ & Calibration iteration window \\
        $\mathcal{H}_k$ & Recent completed iterations preceding iteration $k$ \\
        $\mathcal{A}_{\ell,k}$ & Activated experts in layer $\ell$ at iteration $k$ \\
        $s_{i,k}$ & Tokens routed to expert $i$ at iteration $k$ \\
        $x_i$ & Binary GPU residency decision over $\mathcal{W}$ \\
        $\mathcal{T},\,\mathcal{T}_\ell$ & Model-wide and layer-wise residency targets \\
        $\mathcal{C},\,\mathcal{G}$ & Experts assigned to CPU and GPU execution in the current layer \\
        $m_i$ & Compressed weight size of expert $i$ in bytes \\
        $M_{\mathrm{cache}},\,M_\ell$ & Model-wide and layer-wise GPU cache budgets in bytes \\
        $T_i^{\mathrm{GPU}}(s)$ & GPU execution time for $s$ tokens \\
        $T_i^{\mathrm{CPU}}(s)$ & End-to-end CPU-path time for $s$ tokens \\
        $T_i^{\mathrm{H2D}}$ & H2D expert-weight transfer time \\
        $T_i^{\mathrm{miss}}(s)$ & Estimated execution time on a cache miss \\
        $\Delta T_i(s)$ & Time saved by residency for one access \\
        $S_i$ & Cumulative time saved over $\mathcal{W}$ \\
        $\mathrm{Value}_i$ & Cumulative time saved per byte \\
        $\mathrm{Value}_{i,k}$ & Savings per byte estimated from $\mathcal{H}_k$ \\
        \bottomrule
    \end{tabular}
    \vspace{-0.2in}
\end{table}

\noindent\textbf{Step \circled{5}: Online feedback.}
The scheduler records observed routing demand and execution costs and feeds this information back to the cache planner. The planner uses the updated information to revise residency priorities for subsequent execution, completing the feedback loop shown in Figure~\ref{fig:framework}. Expert ranks and routing decisions remain unchanged throughout inference.

\subsection{Problem Formulation}
\label{sec:formulation}

We consider inference of a low-rank-compressed MoE model on a CPU--GPU platform with limited GPU memory. Let $\mathcal{E}$ denote the set of routed experts across all MoE layers, with $i=(\ell,j)$ identifying expert $j$ in layer $\ell$. Expert ranks and routing assignments are fixed inputs. All expert weights remain available in CPU memory, while a subset can be cached on the GPU. We formulate GPU residency selection over a workload window $\mathcal{W}$ of inference iterations. Table~\ref{tab:notation} summarizes the notation.

At iteration $k\in\mathcal{W}$, expert $i$ receives $s_{i,k}$ tokens, where $s_{i,k}=0$ indicates that it is inactive. The binary decision $x_i=1$ means that its compressed weights, occupying $m_i$ bytes, are retained on the GPU throughout the window. The cache budget $M_{\mathrm{cache}}$ excludes memory reserved for other model state, runtime workspaces, and temporary weight buffers.

For an activated expert processing $s$ tokens, resident GPU execution takes $T_i^{\mathrm{GPU}}(s)$. On a cache miss, the system can either transfer its weights and execute on the GPU, taking $T_i^{\mathrm{H2D}}+T_i^{\mathrm{GPU}}(s)$, or use the CPU path, taking $T_i^{\mathrm{CPU}}(s)$. The CPU-path time includes computation, activation transfers, and synchronization. These times are profiled for the compressed experts on the target hardware. We estimate the cache-miss execution time as
\begin{equation}
    T_i^{\mathrm{miss}}(s)
    =
    \min\left\{
        T_i^{\mathrm{CPU}}(s),\,
        T_i^{\mathrm{H2D}}+T_i^{\mathrm{GPU}}(s)
    \right\}.
    \label{eq:nonresident-cost}
\end{equation}

We then formulate GPU residency selection as
\begin{equation}
\begin{aligned}
    \min_{\{x_i\}}\quad
    &\sum_{k\in\mathcal{W}}
    \sum_{\substack{i\in\mathcal{E}\\s_{i,k}>0}}
    \left[
        x_i T_i^{\mathrm{GPU}}(s_{i,k})
        +(1-x_i)T_i^{\mathrm{miss}}(s_{i,k})
    \right],\\
    \mathrm{s.t.}\quad
    &\sum_{i\in\mathcal{E}}m_i x_i
      \le M_{\mathrm{cache}},\\
    &x_i\in\{0,1\},\qquad \forall i\in\mathcal{E}.
\end{aligned}
    \label{eq:latency-objective}
\end{equation}
The objective minimizes cumulative estimated expert-processing time, while the constraint limits the memory occupied by cached weights. This additive planning model excludes initial cache-loading costs and does not capture overlap or contention among concurrent operations; it is therefore not a direct estimate of end-to-end inference latency.

To derive a cache-selection priority, we define the time saved by GPU residency for one access as
\begin{equation}
\begin{aligned}
    \Delta T_i(s)
    &= T_i^{\mathrm{miss}}(s)-T_i^{\mathrm{GPU}}(s)\\
    &= \min\left\{
        T_i^{\mathrm{H2D}},\,
        T_i^{\mathrm{CPU}}(s)-T_i^{\mathrm{GPU}}(s)
    \right\}.
\end{aligned}
    \label{eq:per-access-latency-benefit}
\end{equation}
Residency saves the weight-transfer time when fetch-and-GPU is the cheaper nonresident path; otherwise, it saves the difference between CPU-path and GPU execution time. A nonpositive $\Delta T_i(s)$ indicates that resident GPU execution offers no time savings for that access. The cumulative savings and savings per byte are
\begin{equation}
    S_i =
    \sum_{\substack{k\in\mathcal{W}\\s_{i,k}>0}}
    \Delta T_i(s_{i,k}),
    \qquad
    \mathrm{Value}_i = \frac{S_i}{m_i}.
    \label{eq:window-value}
\end{equation}
Activation frequency determines how many accesses contribute to $S_i$, while routed-token load determines the savings at each access. Dividing by $m_i$ accounts for the GPU memory required to obtain these savings.

Since the estimated cost without cached experts is independent of $\{x_i\}$, minimizing Equation~\eqref{eq:latency-objective} is equivalent to maximizing $\sum_{i\in\mathcal{E}}S_i x_i$ under the same memory constraint. This yields a 0--1 knapsack problem. RankServe uses $\mathrm{Value}_i$ as a heuristic priority for residency selection, considering only experts with positive cumulative savings. Calibration traces provide the initial demand estimates, while online routing information supports subsequent cache updates.

\begin{algorithm}[t]
\caption{RankServe residency planning and scheduling}
\label{alg:RankServe}
\small
\begin{algorithmic}[1]
\Require Experts $\mathcal{E}$, calibration trace
$\mathcal{W}_{\mathrm{profile}}$, cache budget $M_{\mathrm{cache}}$

\Statex \textbf{Offline planning}
\State Profile expert sizes and CPU, GPU, and H2D times
\State Compute $\{S_i,\mathrm{Value}_i\}$ from
$\mathcal{W}_{\mathrm{profile}}$ using Eq.~\eqref{eq:window-value}
\State Select initial target $\mathcal{T}$ and allocate layer capacities $\{M_\ell\}$

\Statex \textbf{Online decoding}
\For{each decode iteration $k$ and MoE layer $\ell$}
    \State Update $\mathrm{Value}_{i,k}$ from recent routing history
    \State Select and apply target $\mathcal{T}_\ell$ within $M_\ell$
    \State Obtain activated experts $\mathcal{A}_{\ell,k}$ and token counts
    \State $\mathcal{G}\gets\emptyset$; $\mathcal{C}\gets\emptyset$
    \ForAll{$i\in\mathcal{A}_{\ell,k}$}
        \If{$i$ is GPU-resident}
            \State Add $i$ to $\mathcal{G}$
        \ElsIf{$T_i^{\mathrm{CPU}}(s_{i,k})
            \le T_i^{\mathrm{H2D}}+T_i^{\mathrm{GPU}}(s_{i,k})$}
            \State Add $i$ to $\mathcal{C}$
        \Else
            \State Fetch $i$ into temporary GPU storage; add $i$ to $\mathcal{G}$
        \EndIf
    \EndFor
    \State Execute $\mathcal{C}$ on CPU and $\mathcal{G}$ via two-stage grouped GEMMs
    \State Combine expert outputs using routing weights
    \State Record token counts for subsequent residency updates
\EndFor
\end{algorithmic}
\end{algorithm}

\section{RankServe Design and Implementation}
\label{sec:design}

\subsection{System Overview}
\label{sec:design-overview}

Algorithm~\ref{alg:RankServe} summarizes how RankServe coordinates residency planning and expert execution through three components. First, the offline planner combines profiled costs with calibration routing traces to select an initial residency target and allocate GPU cache capacity across layers. Second, the online scheduler updates residency priorities using recent routing history and selects between CPU execution and fetch-and-GPU execution for cache misses based on current token loads. Third, the variable-rank GPU runtime uses Triton to organize heterogeneous expert tasks and CUTLASS grouped GEMMs to execute their low-rank factors without padding them to a common rank. Each layer maintains a fixed-capacity resident cache, while temporarily fetched weights use separate storage.

\subsection{Profile-Guided Residency Target}
\label{sec:initial-placement}

\begin{figure}[t]
    \centering
    \includegraphics[width=\columnwidth]{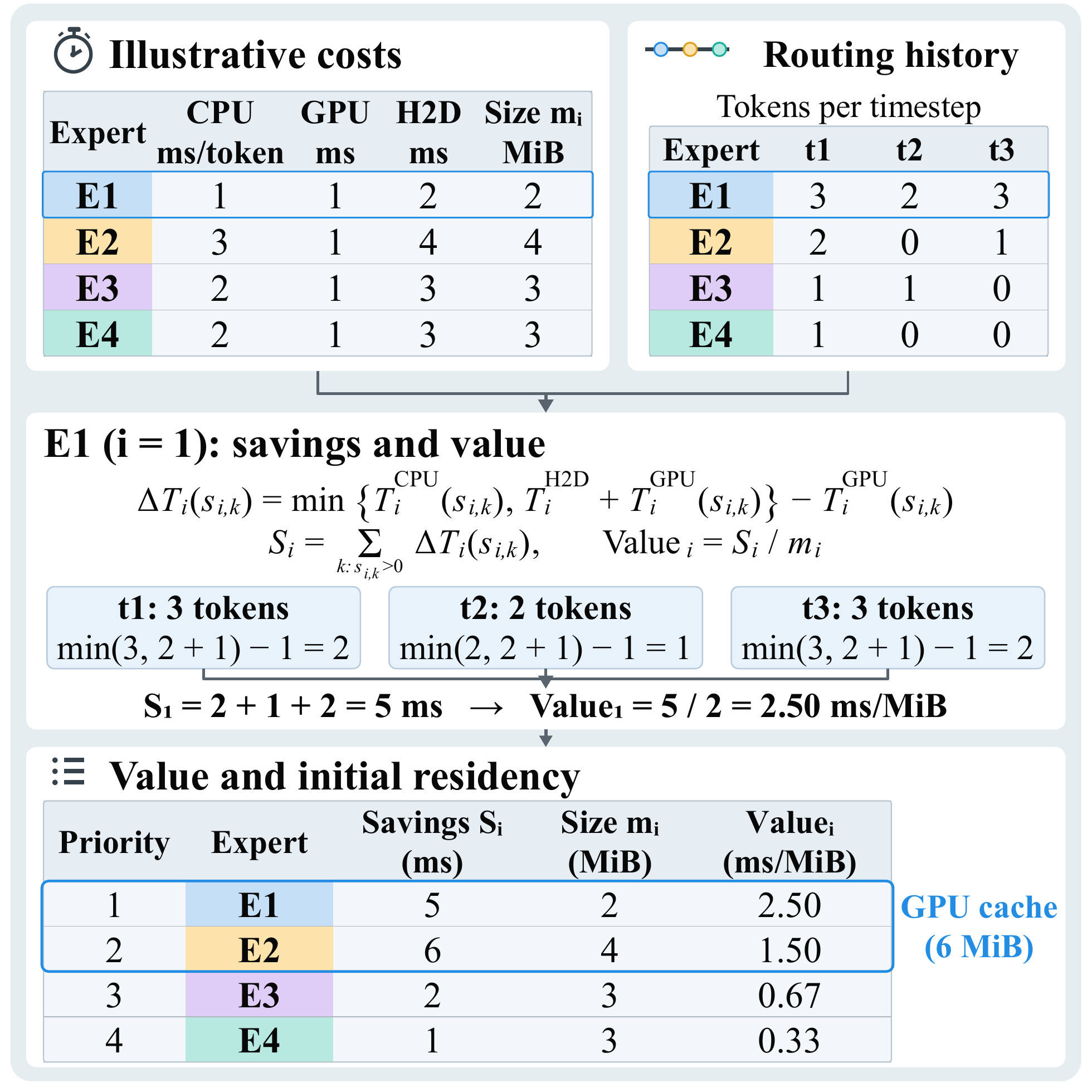}
    \vspace{-0.2in}
    \caption{Profile-guided residency planning. 
    }
    \vspace{-0.2in}
    \label{fig:planner}
\end{figure}

RankServe constructs an initial GPU residency target by combining profiled expert costs with routing demand from calibration inputs. Figure~\ref{fig:planner} illustrates how these inputs determine residency priorities and the resulting cache selection.\lastlinemaybreak 

\noindent\textbf{Cost and demand profiling.}
Before deployment, Rank-Serve records each expert's compressed weight size $m_i$ and profiles its H2D weight-transfer time $T_i^{\mathrm{H2D}}$, GPU execution time $T_i^{\mathrm{GPU}}(s)$, and end-to-end CPU-path time $T_i^{\mathrm{CPU}}(s)$ across routed-token loads. Profiling uses the same weight layout and execution paths as inference. RankServe also collects a calibration trace $\mathcal{W}_{\mathrm{profile}}$ recording the token count $s_{i,k}$ assigned to each expert at each iteration. This trace captures both activation frequency and the workload of each activation.

\noindent\textbf{Residency priority.}
For each expert, RankServe evaluates the per-access savings in Equation~\eqref{eq:per-access-latency-benefit} using the profiled costs and observed token loads. It then computes $S_i$ and $\mathrm{Value}_i$ from Equation~\eqref{eq:window-value} with $\mathcal{W}=\mathcal{W}_{\mathrm{profile}}$. The resulting priority measures cumulative estimated latency savings per byte of GPU memory. It favors experts whose repeated GPU execution avoids substantial nonresident costs relative to their weight sizes.

\begin{figure}[!t]
    \centering
    \includegraphics[width=\columnwidth]{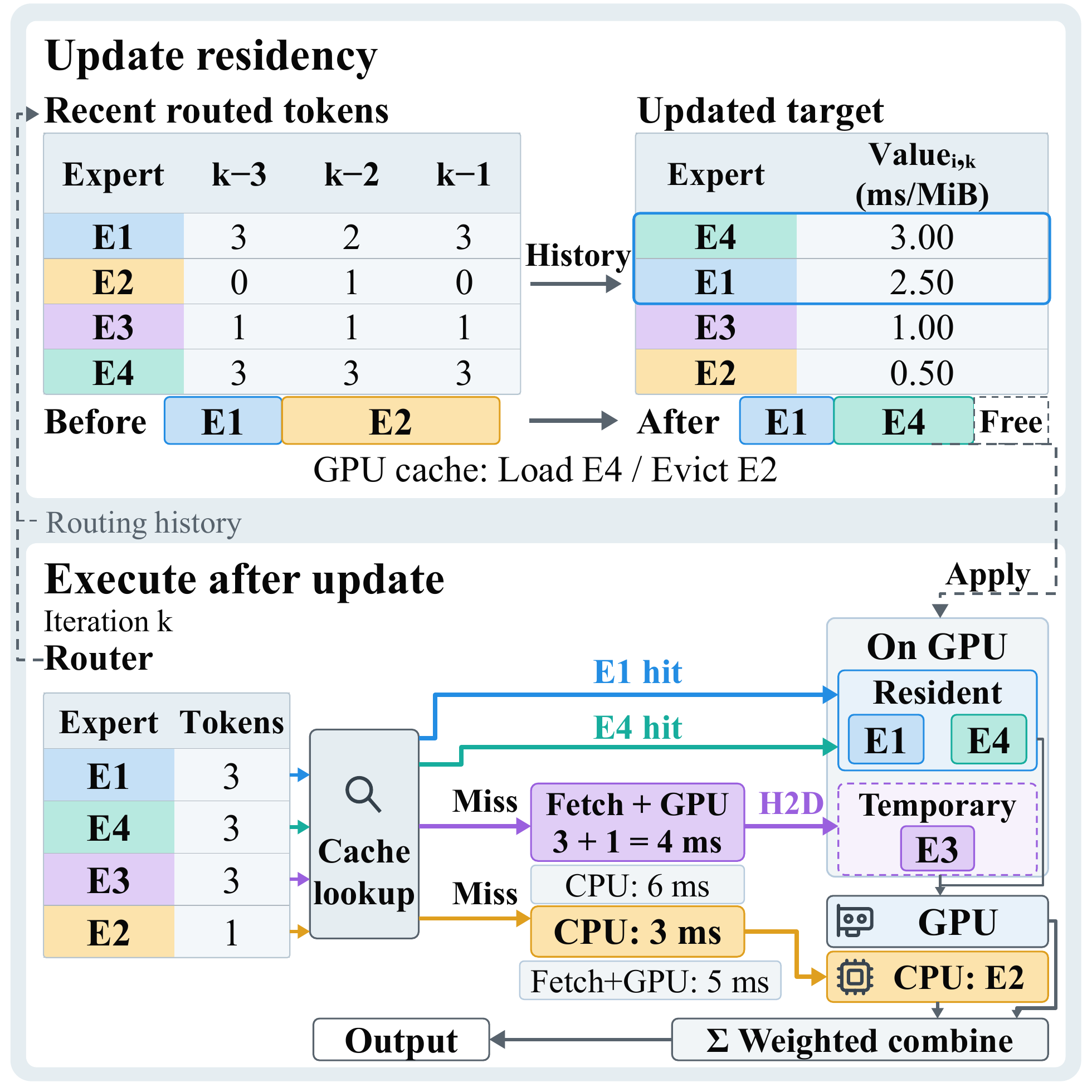}
    \caption{Online decode scheduling. 
    }
    \vspace{-0.15in}
    \label{fig:scheduler}
\end{figure}

\noindent\textbf{Target selection.}
RankServe sorts experts with positive cumulative savings by decreasing $\mathrm{Value}_i$ and adds each expert that fits within the remaining cache budget. It also considers the single feasible expert with the largest $S_i$ and retains whichever selection provides greater total estimated savings. This heuristic accounts for both residency benefit and memory usage without solving the knapsack problem exactly. In Figure~\ref{fig:planner}, it selects E1 and E2 under a 6\,MiB cache budget.

\noindent\textbf{Cache initialization.}
The selected experts determine the minimum cache capacity required by each MoE layer. RankServe distributes the remaining budget among layers in proportion to their unselected expert weight sizes and aligns the capacities to the GPU weight layout while respecting the total budget. The residency target identifies experts to retain, but their weights are loaded on first access. If complete residency-value profiles are unavailable, RankServe instead allocates layer capacities in proportion to their compressed expert weight sizes and relies on online admission and replacement.

\subsection{Online Decode Scheduling}
\label{sec:online-scheduling}

The initial residency target reflects the calibration workload, whereas expert demand changes during decoding. RankServe adapts by updating residency priorities from recent routing history and selecting execution paths using current routed-token loads. Figure~\ref{fig:scheduler} illustrates these two decisions: recent demand guides cache replacement, while current routing determines how each activated expert executes.

\noindent\textbf{Online residency update.}
RankServe records the token counts assigned to each expert during decoding. Before iteration $k$, let $\mathcal{H}_k$ denote a window of recent completed iterations. Using the profiled per-access savings from Equation~\eqref{eq:per-access-latency-benefit}, it updates each expert's residency priority as
\begin{equation}
    \mathrm{Value}_{i,k}
    =
    \frac{1}{m_i}
    \sum_{\substack{t\in\mathcal{H}_k\\s_{i,t}>0}}
    \Delta T_i(s_{i,t}),
    \label{eq:online-residency-value}
\end{equation}
where $t$ indexes historical iterations and $s_{i,t}$ is the observed routed-token count. This score captures changes in both activation frequency and workload per activation, while accounting for expert weight size.

The cache planner uses these updated priorities to revise the residency target within each layer's allocated capacity. The scheduler applies the target through admission and replacement, retaining higher-priority experts and releasing lower-priority cached copies when space is needed. All expert weights remain available in CPU memory.

\begin{figure}[!t]
    \centering
    \includegraphics[width=0.95\columnwidth]{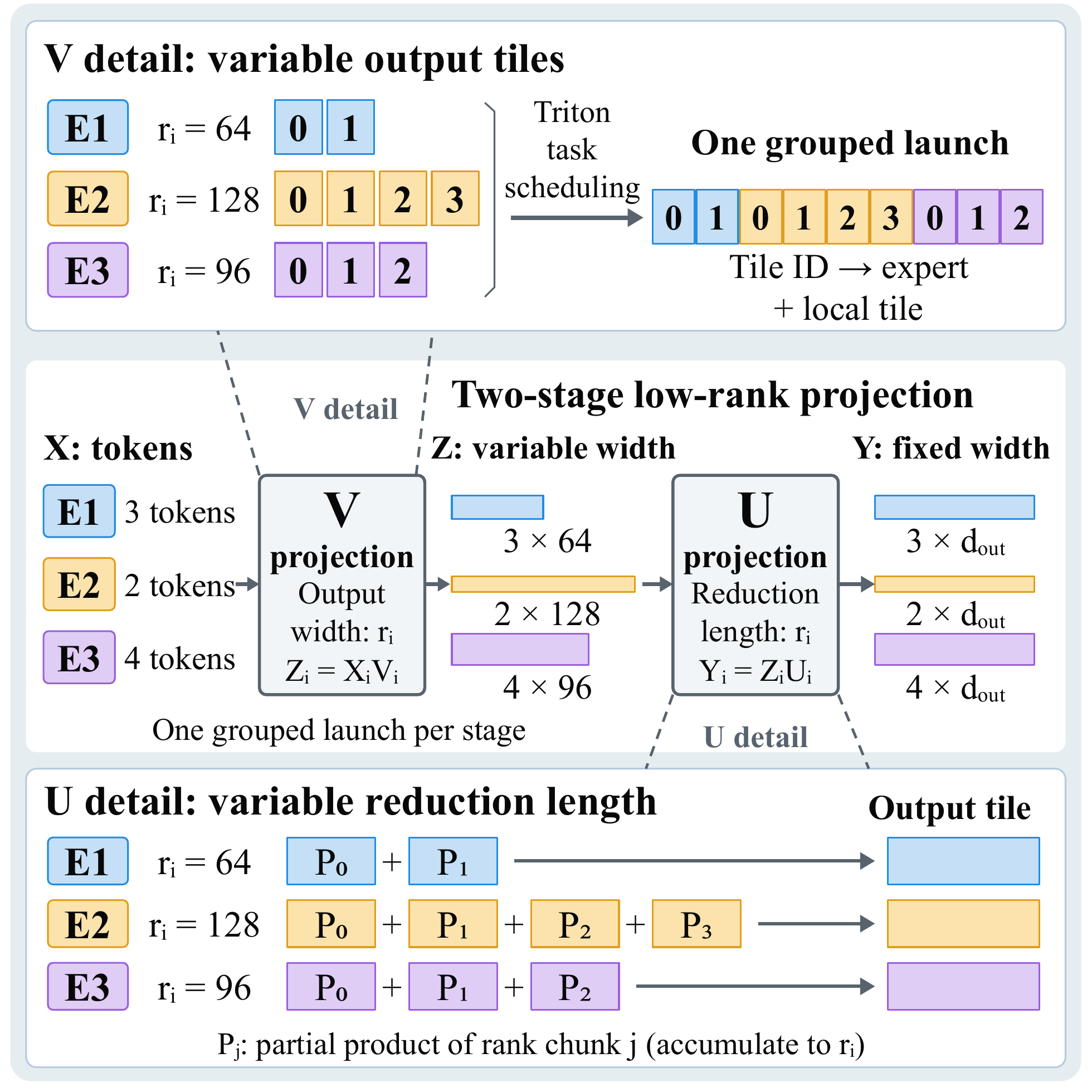}
      \caption{Variable-rank GPU runtime. Triton organizes heterogeneous expert tasks for CUTLASS grouped execution. V projection varies in output width, while U projection varies in reduction length, with one grouped launch per stage.}
    \label{fig:runtime}
    \vspace{-0.1in}
\end{figure}

\noindent\textbf{Execution-path selection.}
After routing at iteration $k$, RankServe checks the activated experts against the current cache state. Resident experts execute directly on the GPU. For a nonresident expert $i$, the scheduler compares the CPU-path time with the time required to transfer its weights and execute on the GPU:
\begin{equation}
    \mathrm{Path}_{i,k}
    =
    \begin{cases}
        \mathrm{CPU},
        & T_i^{\mathrm{CPU}}(s_{i,k})
        \le
        T_i^{\mathrm{H2D}}+T_i^{\mathrm{GPU}}(s_{i,k}),\\
        \mathrm{Fetch\text{-}and\text{-}GPU},
        & \text{otherwise}.
    \end{cases}
    \label{eq:online-execution-path}
\end{equation}
Here, $\mathrm{Path}_{i,k}$ denotes the selected execution path for that cache miss. The CPU path transfers the required activations while keeping weights in CPU memory; the fetch-and-GPU path transfers the compressed weights. Since the execution times depend on the current token count, an expert can use different paths across iterations.

\noindent\textbf{Execution and cache admission.}
Fetching an expert for GPU execution does not automatically admit it to the resident cache. Experts outside the residency target can execute from temporary GPU weight storage, allowing the scheduler to use the GPU without displacing more valuable cached experts. Conversely, CPU execution leaves the expert weights in host memory. In Figure~\ref{fig:scheduler}, E1 and E4 are cache hits, E3 is fetched for GPU execution, and E2 executes on the CPU. The resulting expert outputs are combined using the routing weights, and the observed token counts are recorded for subsequent residency updates.

\subsection{Variable-Rank GPU Runtime}
\label{sec:variable-rank-runtime}

Experts dispatched to the GPU can have different retained ranks and routed-token loads, producing heterogeneous matrix shapes. RankServe uses Triton to organize and schedule these tasks and CUTLASS grouped GEMM kernels to execute them. Figure~\ref{fig:runtime} illustrates how the runtime handles rank variation in the two stages of a low-rank projection.

\noindent\textbf{Two-stage low-rank execution.}
Let $X_i\in\mathbb{R}^{s_i\times d_{\mathrm{in}}}$ contain the $s_i$ tokens assigned to expert $i$ in the current iteration. For row-wise token processing, we express each expert's factorization $W_i \approx A_iB_i$ using transposed factors $V_i=B_i^{\mathsf T}$ and $U_i=A_i^{\mathsf T}$. The projection is computed in two stages:
\begin{equation}
    Z_i=X_iV_i,
    \qquad
    Y_i=Z_iU_i,
    \label{eq:runtime-projections}
\end{equation}
where $V_i\in\mathbb{R}^{d_{\mathrm{in}}\times r_i}$ and $U_i\in\mathbb{R}^{r_i\times d_{\mathrm{out}}}$. The intermediate $Z_i$ has shape $s_i\times r_i$, while the output $Y_i$ has shape $s_i\times d_{\mathrm{out}}$. The retained rank $r_i$ therefore determines the output width of the V projection and the reduction length of the U projection.

\noindent\textbf{Task organization and grouped execution.}
RankServe stores compressed factors at their assigned ranks, with shape and weight-offset metadata. After vLLM groups tokens by expert, Triton organizes their matrix operations for grouped execution. For the $V$ projection, output widths vary with $r_i$. Triton packs valid output-tile tasks across experts, mapping each to its expert and local tile index (top of Figure~\ref{fig:runtime}). For the $U$ projection, output width is fixed but reduction length varies: each tile accumulates over rank chunks up to $r_i$, with boundary handling for partial chunks. CUTLASS executes each stage in one grouped launch, avoiding per-rank launches and the storage and computation overhead of padding experts to the maximum rank.


\begin{figure*}[!t]
    \centering
    \includegraphics[width=0.95\textwidth]{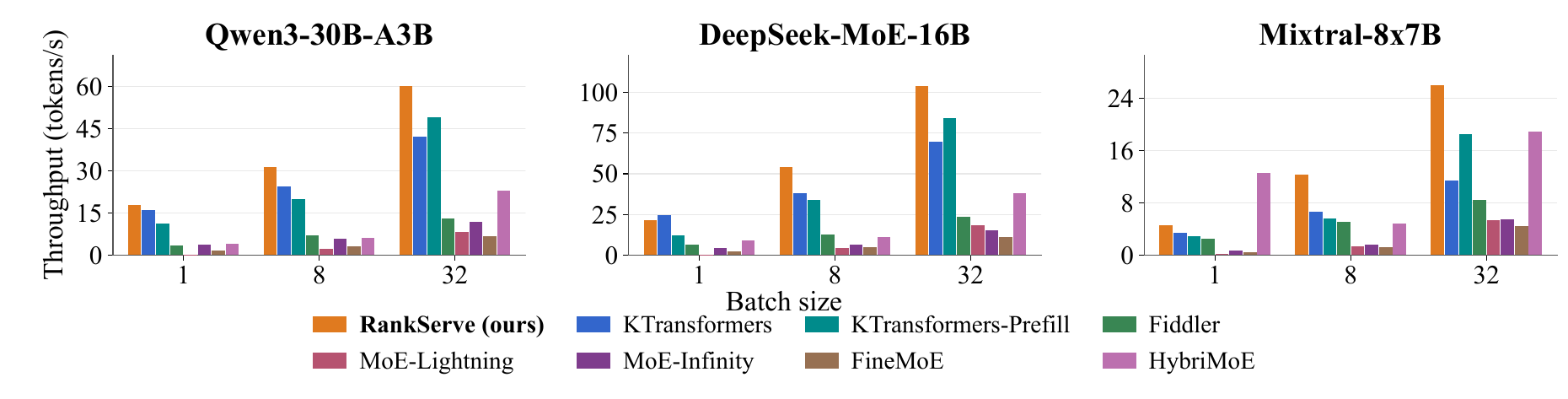}
    \vspace{-0.25in}
    \caption{Throughput comparison across three MoE models at batch sizes 1, 8, and 32, using 128-token inputs and outputs. GPU weight budgets are 12 GiB for Qwen3-30B-A3B and DeepSeek-MoE-16B and 24 GiB for Mixtral-8$\times$7B.}
    \label{fig:throughput}
    \vspace{-0.15in}
\end{figure*}

\begin{figure}[t]
    \centering
    \includegraphics[width=0.95\columnwidth]{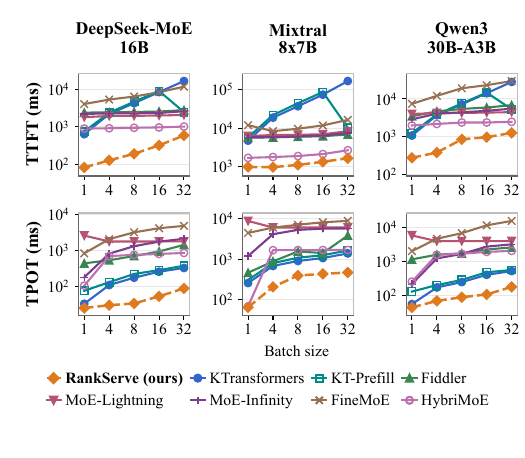}
    \vspace{-0.35in}
    \caption{Inference latency across batch sizes using 256 prompts from ShareGPT with 128 output tokens per request.
    }
    \label{fig:main_results}
    \vspace{-0.1in}
\end{figure}

\noindent\textbf{Transfer and execution coordination.}
Compressed expert weights remain in pinned CPU memory. Asynchronous H2D transfers place GPU-dispatched experts in either the resident cache or temporary weight storage, and CUDA events ensure that dependent computation starts only after the required weights arrive. Independent transfers and GPU computation can overlap. For experts dispatched to the CPU, computation begins after the input activation transfer completes, and the outputs return to the GPU before the routing-weighted combination. Both resident and temporarily fetched GPU experts use the same variable-rank execution path.

\section{Evaluation}
\label{sec:evaluation}

\subsection{Experimental Setup}
\label{sec:eval-setup}

\noindent\textbf{Models and Baselines.}
We evaluate DeepSeek-MoE-16B, Mixtral-8x7B, and Qwen3-30B-A3B. Unless otherwise specified, RankServe uses RFID-MoE~\cite{mi2026effective} with a 40\% compression ratio and fixed expert ranks during inference. Baselines use the corresponding models without low-rank factorization, retaining their respective weight representations and quantization schemes. We compare against six systems: KTransformers~\cite{chen2025ktransformers}, Fiddler~\cite{kamahori2025fiddler}, MoE-Lightning~\cite{cao2025moe}, MoE-Infinity~\cite{xue2024moe}, FineMoE~\cite{yu2026taming}, and HybriMoE~\cite{zhong2025hybrimoe}. For KTransformers, we evaluate both CPU execution of routed experts during prefill and its open-source full-layer GPU prefill path (KT-Prefill).%
\looseness=1 

\noindent\textbf{Testbed and Workloads.} 
We use a single NVIDIA RTX A6000 GPU, a 32-core AMD Ryzen Threadripper PRO 5975WX CPU, and 256\,GB of DDR4-2666 memory. RankServe builds on vLLM~\cite{kwon2023efficient}, with the GPU expert-weight budget specified per experiment.
%
We use 256 prompts from ShareGPT and generate 128 output tokens per request. The main experiments vary the batch size over $\{1,4,8,16,32\}$ to evaluate performance under different amounts of concurrent expert demand. All systems process the same prompts and generate the same number of output tokens.

\noindent\textbf{Evaluation Metrics.}
We report time to first token (TTFT), time per output token (TPOT), and throughput. TTFT spans the inference API call to the first output token. TPOT is the time between the first and last output tokens divided by the number of output tokens minus one. Throughput is the total number of generated tokens across the batch divided by end-to-end inference time, including prefill and decode. Each configuration uses one warm-up run and two measured runs; we report their arithmetic mean.


\subsection{Accuracy with Heterogeneous Rank Allocation}
\label{sec:eval-rank-accuracy}

\begin{table}[!t]
    \centering
    \caption{Effect of heterogeneous rank allocation on Qwen3-30B-A3B at 60\% compression. Bold values indicate the better result within each method.}
    \vspace{-0.1in}
    \label{tab:rank_allocation}
    \small
    \setlength{\tabcolsep}{3pt}
    \renewcommand{\arraystretch}{1.05}
    \begin{tabular}{@{}lcccc@{}}
        \toprule
        & \multicolumn{3}{c}{PPL $\downarrow$}
        & ACC (\%) $\uparrow$ \\
        \cmidrule(lr){2-4}
        Method & WikiText-2 & PTB & C4 & Avg. \\
        \midrule
        SVD-LLM & 16.79 & 33.27 & 37.22 & 50.16 \\
        +Hetero. Rank & \textbf{10.07} & \textbf{21.59} & \textbf{24.90} & \textbf{50.77} \\
        \midrule
        D$^{2}$-MoE & 16.87 & 33.92 & 37.70 & 49.73 \\
        +Hetero. Rank & \textbf{10.05} & \textbf{22.19} & \textbf{24.62} & \textbf{50.99} \\
        \midrule
        MoLAE & 21.20 & 37.38 & 36.25 & 53.94 \\
        +Hetero. Rank & \textbf{17.57} & \textbf{32.40} & \textbf{32.38} & \textbf{54.17} \\
        \midrule
        MoBE & 13.96 & 24.93 & 24.40 & 61.33 \\
        +Hetero. Rank & \textbf{10.43} & \textbf{18.25} & \textbf{19.15} & \textbf{62.59} \\
        \bottomrule
    \end{tabular}
    \vspace{-0.2in}
\end{table}

We evaluate heterogeneous rank allocation on Qwen3-30B-A3B at a 60\% compression ratio using four low-rank compression methods. Table~\ref{tab:rank_allocation} reports perplexity on WikiText-2 \cite{merity2016pointer}, PTB \cite{marcus1993building}, and C4 \cite{raffel2020exploring}, together with average accuracy across OpenBookQA \cite{mihaylov2018can}, ARC-Easy \cite{clark2018think}, WinoGrande \cite{sakaguchi2021winogrande}, HellaSwag \cite{zellers2019hellaswag}, PIQA \cite{bisk2020piqa}, and MathQA \cite{amini2019mathqa}. Heterogeneous rank allocation reduces perplexity on all three datasets for every method and improves average accuracy by 0.23--1.26 percentage points. These results support the use of non-uniform ranks to preserve model quality and motivate efficient inference support for variable-rank experts.

\subsection{End-to-End Performance}
\label{sec:eval-overall}

\noindent\textbf{Throughput.}
Figure~\ref{fig:throughput} compares throughput across three models and batch sizes. RankServe achieves the highest throughput on Qwen3-30B-A3B and DeepSeek-MoE-16B at all tested batch sizes, reaching 61.5 and 104.2 tokens/s at batch size 32, respectively. On Mixtral-8x7B, RankServe achieves 12.5 and 25 tokens/s at batch sizes 8 and 32, respectively, but trails HybriMoE at batch size 1. These results demonstrate improved throughput across most evaluated configurations, with gains varying by model and batch size.

\begin{figure}[t]
    \centering
    \includegraphics[width=\columnwidth]{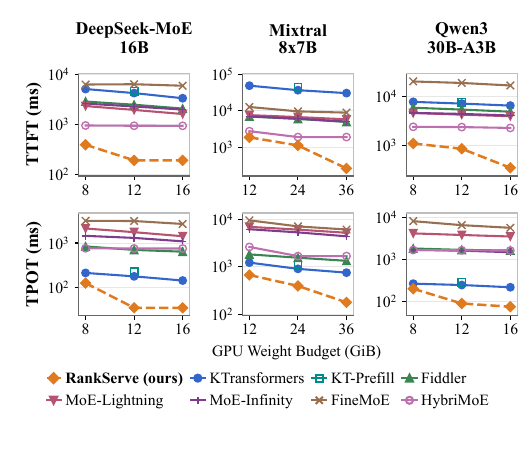}
    \vspace{-0.5in}
      \caption{Performance under varying GPU weight budgets at batch size 8. 
      }
    \label{fig:memory-budget}
    \vspace{-0.2in}
\end{figure}

\begin{figure}[t]
    \centering
    \includegraphics[width=0.9\columnwidth]
        {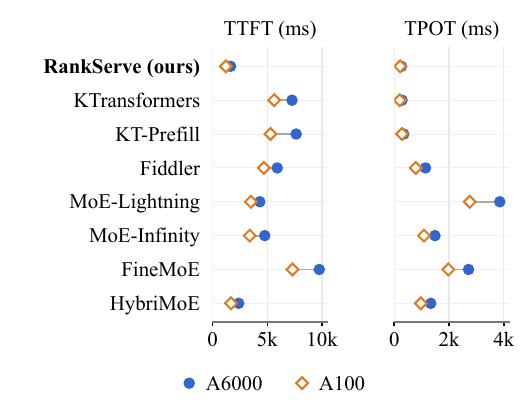}
        \vspace{-0.2in}
    \caption{Latency on A6000 and A100 platforms:
    Qwen3-30B-A3B, batch size 8, 128-token inputs and
    outputs, and a 12\,GiB GPU weight budget.}
    \vspace{-0.2in}
    \label{fig:hardware}
\end{figure}

\noindent\textbf{Latency.}
Figure~\ref{fig:main_results} compares TTFT and TPOT at batch sizes 1--32 using the same GPU weight budgets as Figure~\ref{fig:throughput}. Speedups are measured against the fastest baseline for each configuration and metric. Across the five batch sizes, RankServe achieves geometric-mean TTFT speedups of 2.29$\times$, 1.69$\times$, and 2.12$\times$ for DeepSeek-MoE-16B, Mixtral-8x7B, and Qwen3-30B-A3B, respectively; corresponding TPOT speedups are 2.35$\times$, 1.71$\times$, and 2.48$\times$.
HybriMoE achieves the lowest baseline TTFT in most configurations, while KTransformers is generally more competitive in TPOT. RankServe accelerates both phases through expert residency management, workload-dependent CPU--GPU execution, and grouped variable rank computation. Gains vary across models and batch sizes with expert sizes, routing demand, and execution workloads. Ablations evaluate each component's contribution.



\begin{figure}[t]
    \centering
\includegraphics[width=\columnwidth]{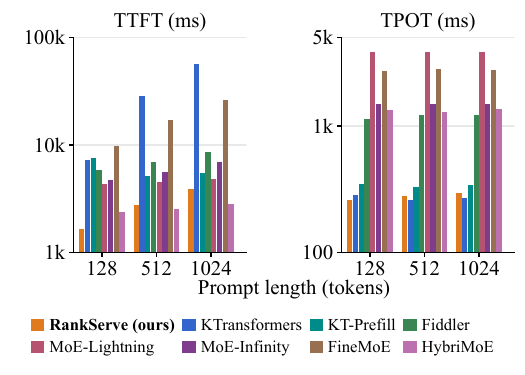}
\vspace{-0.3in}
    \caption{Latency under different prompt length on Qwen3-30B-A3B with batch size 8, 128 output tokens, and a 12\,GiB GPU weight budget. Both panels use logarithmic scales.}
    \vspace{-0.2in}
    \label{fig:prompt-length}
\end{figure}

\begin{figure}[!t]
    \centering
    \includegraphics[width=0.9\columnwidth]{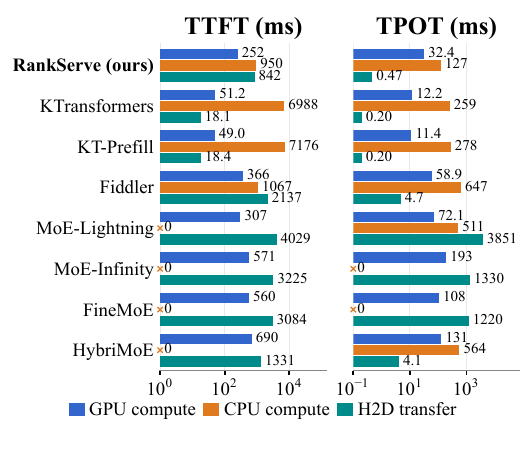}
    \vspace{-0.3in}
    \caption{GPU computation, CPU computation, and H2D transfer times on Qwen3-30B-A3B with batch size 8 and a 12\,GiB GPU weight budget. Prefill times are measured per batch and decode times per generation step.}
    \label{fig:runtime-breakdown}
    \vspace{-0.05in}
\end{figure}

At batch size of 8, RankServe achieves the lowest TTFT and TPOT across all evaluated GPU weight budgets (Figure~\ref{fig:memory-budget}), with speedups of $1.48$--$2.31\times$ and $1.32$--$3.15\times$, respectively, over the best measured baseline per configuration. Larger budgets generally reduce latency: Mixtral and Qwen3 continue to improve, while DeepSeek shows diminishing returns. Its TTFT and TPOT reach approximately 191\,ms and 34\,ms at 12\,GiB, with little change at 16\,GiB. These results indicate a favorable latency--memory trade-off under the evaluated compression setting.


\subsection{Performance Under Different Hardware}
\label{sec:eval-hardware}

Figure~\ref{fig:hardware} compares the A6000 and A100 platforms, both using PCIe 4.0 $\times$16 links. The moderate latency reductions are consistent with CPU and transfer bottlenecks that limit the benefits of higher GPU compute capability. RankServe achieves the lowest TTFT on both platforms and remains competitive in TPOT.

\subsection{Latency Under Different Prompt Length}
\label{sec:eval-prompt-length}

Figure~\ref{fig:prompt-length} compares TTFT and TPOT under different prompt lengths. Longer prompts increase RankServe's TTFT, while its TPOT changes relatively little. RankServe achieves the lowest TTFT at 128 tokens and remains competitive at 512 tokens.


\subsection{Runtime Breakdown Analysis}
\label{sec:eval-analysis}

Figure~\ref{fig:runtime-breakdown} breaks down GPU computation, CPU computation, and H2D transfer times. During prefill, KTransformers spends 6988\,ms on CPU computation, while MoE-Infinity and FineMoE incur over 3000\,ms of H2D transfers. RankServe distributes work across CPU and GPU execution, with CPU computation and H2D transfer times of 950\,ms and 842\,ms, respectively. During decode, RankServe reduces CPU computation to 127\,ms, compared with 259\,ms for KTransformers and 564\,ms for HybriMoE, while keeping H2D transfer time at 0.47\,ms. These results suggest that RankServe achieves a more balanced use of computation and communication paths, reducing reliance on CPU execution or expert-weight transfers. The measured components can overlap and therefore do not sum to end-to-end latency.

\subsection{Ablation Studies}
\label{sec:eval-ablation}

\begin{table}[t]
    \centering
    \caption{Ablation study of RankServe on Qwen3-30B-A3B with BF16 heterogeneous-rank factors retaining 50\% of expert parameters, a 12 GiB GPU weight budget, batch size 8, and 128-token inputs and outputs.}
    \vspace{-0.1in}
    \label{tab:ablation}
    \small
    \setlength{\tabcolsep}{4pt}
    \renewcommand{\arraystretch}{1.1}
    \begin{tabular*}{\columnwidth}{@{\extracolsep{\fill}}lrr@{}}
        \toprule
        Configuration & TTFT (ms) & TPOT (ms) \\
        \midrule
        All optimizations disabled & 1965 & 396.90 \\
        w/o Prefill Placement & 1745 & 252.43 \\
        w/o Online Feedback & 1531 & 345.96 \\
        w/o Variable-Rank GPU Runtime & 1795 & 350.71 \\
        \textbf{Full RankServe} & \textbf{1530} & \textbf{234.17} \\
        \bottomrule
    \end{tabular*}
    \vspace{-0.2in}
\end{table}

\noindent\textbf{Effects of different optimization techniques.} We evaluate the contributions of prefill placement, online feedback, and the variable-rank GPU runtime by disabling each component individually and all three together. As shown in Table~\ref{tab:ablation}, full RankServe reduces TTFT and TPOT by 22.1\% and 41.0\%, respectively, compared with disabling all optimizations. Removing the variable-rank GPU runtime causes the largest overall degradation among the individual ablations, increasing TTFT by 17.3\% and TPOT by 49.8\% relative to full RankServe. In addition, disabling online feedback causes a substantial TPOT degradation of 47.7\% with negligible effect on TTFT, while disabling prefill placement increases TTFT by 14.1\% and TPOT by 7.8\%.

\noindent\textbf{Effects of linear activations.}
To facilitate system optimization, we replace the SiLU
activations in MoBE and RFID with linear activations.
Table~\ref{tab:linear_activation} shows that this modification
preserves model quality: both methods achieve lower perplexity
on all three datasets and higher average accuracy across
the six zero-shot tasks.

\section{Related Work}
\label{sec:related-work}
\paragraph{Low-Rank Compression for MoE Models.}
Low-rank factorization compresses weights~\cite{wang2024model}, while activation-aware methods reduce truncation error~\cite{wang2025svd,wang2025svd2,bai2025erc,wang2025dobi}. MoE compression additionally addresses heterogeneous expert sensitivity and cross-expert redundancy~\cite{chen2026mobe,li2026sub,xu2026camera,lai2026rs}. MoE-SVD~\cite{li2025structured} selectively decomposes experts and shares low-rank components; D$^2$-MoE~\cite{gu2025delta} compresses expert-specific deltas from a shared base. MoLAE~\cite{liu2025molae} and MoBE~\cite{chen2026mobe} share projections or bases across experts. RFID-MoE~\cite{mi2026effective} assigns heterogeneous ranks using routing frequency and effective-rank-based information density, producing unequal memory, transfer, and computation costs. RankServe coordinates the residency and CPU--GPU execution of these variable-rank experts under limited GPU memory.


\paragraph{Memory-Constrained MoE Inference and Expert Offloading.}
Expert offloading reduces GPU memory by storing experts on the CPU and loading them on demand~\cite{liu2026survey}. 
To reduce loading stalls, Mixtral-Offloading~\cite{eliseev2023fast} caches and prefetches experts, while MoE-Infinity~\cite{xue2024moe}, ProMoE~\cite{song2024promoe}, and FineMoE~\cite{yu2026taming} predict expert demand from activation history, intermediate states, or historical expert maps.
However, these systems do not consider rank-dependent memory and nonresident execution costs.
RankServe combines these costs with access demand to prioritize GPU residency by expected latency reduction per byte.

\begin{table}[t]
\centering
\caption{SiLU versus linear activations.
Accuracy is averaged over six zero-shot tasks.}
\vspace{-0.1in}
\label{tab:linear_activation}
\fontsize{9}{10}\selectfont
\setlength{\tabcolsep}{2pt}
\begin{tabular*}{\columnwidth}{
@{\extracolsep{\fill}}llrrrr@{}
}
\toprule
& & \multicolumn{3}{c}{PPL $\downarrow$}
& ACC (\%) $\uparrow$ \\
\cmidrule(lr){3-5}
Method & Act. & WikiText-2 & PTB & C4 & Avg. \\
\midrule
MoBE\cite{chen2026mobe} & SiLU
& 13.96 & 24.93 & 24.40 & 61.33 \\
& Linear
& \textbf{13.23} & \textbf{23.35}
& \textbf{23.45} & \textbf{62.76} \\
\midrule
RFID\cite{mi2026effective} & SiLU
& 9.57 & 16.92 & 17.96 & 64.17 \\
& Linear
& \textbf{9.50} & \textbf{16.46}
& \textbf{17.47} & \textbf{65.71} \\
\bottomrule
\end{tabular*}
\vspace{-0.1in}
\end{table}

\paragraph{CPU--GPU Collaborative LLM Inference.}
Beyond storing offloaded weights, CPUs can also execute model components to reduce transfers~\cite{liu2026survey}. 
MoE-Lightning~\cite{cao2025moe} executes decode-stage attention on the CPU to avoid KV-cache transfers while overlapping CPU/GPU computation and expert loading. 
For expert execution, Fiddler~\cite{kamahori2025fiddler} selects between CPU execution and GPU loading based on token counts and device latencies, while HOBBIT~\cite{tang2024hobbit} reduces cache-miss loading costs through mixed-precision loading, prefetching, and caching.
HybriMoE~\cite{zhong2025hybrimoe} balances expert computation across CPUs and GPUs, DAOP~\cite{zhang2025daop} pre-computes predicted CPU-resident experts, and KTransformers~\cite{chen2025ktransformers} accelerates CPU experts with optimized kernels and asynchronous scheduling.
In contrast, RankServe targets variable-rank experts, using rank-dependent execution costs and routed-token loads to select execution paths, with a GPU runtime supporting heterogeneous factor dimensions.

\section{Conclusion}
\label{sec:conclusion}
This paper presents RankServe, a CPU--GPU scheduling system for low-rank MoE inference under limited GPU memory. RankServe combines rank-dependent costs with routing demand to prioritize GPU residency by estimated latency savings per byte relative to the cheaper nonresident path. It updates residency priorities from recent routing history and uses current token loads to choose between CPU execution and weight transfer followed by GPU execution. Its variable-rank runtime executes heterogeneous factors through grouped GEMMs without common-rank padding. Built on vLLM, RankServe improves latency and throughput over existing MoE serving systems across diverse models, batch sizes, and GPU memory budgets.

\bibliographystyle{unsrtnat}
\bibliography{refs}

@inproceedings{wang2025svd,
  title={Svd-llm: Truncation-aware singular value decomposition for large language model compression},
  author={Wang, Xin and Zheng, Yu and Wan, Zhongwei and Zhang, Mi},
  booktitle={International Conference on Learning Representations},
  volume={2025},
  pages={19299--19319},
  year={2025}
}

@article{li2025structured,
  title={Structured mixture-of-experts LLMs compression via singular value decomposition},
  author={Li, Wei and Li, Lujun and Huang, You-Liang and Lee, Mark G and Sun, Shengjie and Xue, Wei and Guo, Yike},
  year={2025}
}

@article{gu2025delta,
  title={Delta decompression for moe-based llms compression},
  author={Gu, Hao and Li, Wei and Li, Lujun and Zhu, Qiyuan and Lee, Mark and Sun, Shengjie and Xue, Wei and Guo, Yike},
  journal={arXiv preprint arXiv:2502.17298},
  year={2025}
}

@article{yang2026freetoken,
  title={FreeToken: Efficient Edge-Native MoE Serving with Bandwidth-Adaptive Execution},
  author={Yang, Shuo and Fan, Xiaoze and Pan, Melissa and Xi, Haocheng and Wang, Zhe and Sun, Shanlin and Keutzer, Kurt and Han, Song and Zaharia, Matei and Xu, Chenfeng and others},
  journal={arXiv preprint arXiv:2608.16157},
  year={2026}
}

@article{li2026moe,
  title={MoE-SpAc: Efficient MoE Inference Based on Speculative Activation Utility in Heterogeneous Edge Scenarios},
  author={Li, Shuhuai and Lin, Jianghao and Ge, Dongdong and Ye, Yinyu},
  journal={arXiv preprint arXiv:2603.09983},
  year={2026}
}

@inproceedings{tang2026moe,
  title={MoE-APEX: An Efficient MoE Inference System with Adaptive Precision Expert Offloading},
  author={Tang, Peng and Liu, Jiacheng and Hou, Xiaofeng and Pu, Yifei and Wang, Jing and Heng, Pheng-Ann and Li, Chao and Guo, Minyi},
  booktitle={Proceedings of the 31st ACM International Conference on Architectural Support for Programming Languages and Operating Systems, Volume 2},
  pages={1185--1200},
  year={2026}
}

@article{liu2025molae,
  title={Molae: Mixture of latent experts for parameter-efficient language models},
  author={Liu, Zehua and Wu, Han and She, Ruifeng and Fu, Xiaojin and Han, Xiongwei and Zhong, Tao and Yuan, Mingxuan},
  journal={arXiv preprint arXiv:2503.23100},
  year={2025}
}

@article{tang2024hobbit,
  title={Hobbit: A mixed precision expert offloading system for fast moe inference},
  author={Tang, Peng and Liu, Jiacheng and Hou, Xiaofeng and Pu, Yifei and Wang, Jing and Heng, Pheng-Ann and Li, Chao and Guo, Minyi},
  journal={arXiv preprint arXiv:2411.01433},
  year={2024}
}

@article{zhang2026wisp,
  title={WiSP: A Working-Set View of Mixture-of-Experts Serving on Extremely Low-Resource Hardware},
  author={Zhang, Jiamu and Wu, Liang and Darbari, Mayank and Hong, Liangjie},
  journal={arXiv preprint arXiv:2606.21868},
  year={2026}
}

@article{liu2025bandwidth,
  title={Bandwidth-Efficient Adaptive Mixture-of-Experts via Low-Rank Compensation},
  author={Liu, Zhenyu and Liu, Yunzhen and Fan, Zehao and Gagnon, Garrett and Hou, Yayue and Wu, Nan and Kang, Yangwook and Liu, Liu},
  journal={arXiv preprint arXiv:2512.17073},
  year={2025}
}

@inproceedings{yang2026faster,
  title={Faster moe llm inference for extremely large models},
  author={Yang, Haoqi and Shi, Luohe and Li, Qiwei and Li, Zuchao and Wang, Ping and Huang, Hao and Zhao, Hai},
  booktitle={Findings of the Association for Computational Linguistics: ACL 2026},
  pages={43133--43151},
  year={2026}
}

@inproceedings{chen2026mobe,
  title={MoBE: Mixture-of-Basis-Experts for Compressing MoE-based LLMs},
  author={Chen, Xiaodong and Ha, Mingming and Lan, Zhenzhong and Zhang, Jing and Li, Jianguo},
  booktitle={International Conference on Learning Representations},
  volume={2026},
  pages={40010--40028},
  year={2026}
}

@article{mi2026effective,
  title={Effective moe-based llm compression by exploiting heterogeneous inter-group experts routing frequency and information density},
  author={Mi, Zhendong and Chen, Yixiao and Zhao, Pu and Yu, Xiaodong and Wang, Hao and Wang, Yanzhi and Huang, Shaoyi},
  journal={arXiv preprint arXiv:2602.09316},
  year={2026}
}

@article{eliseev2023fast,
  title={Fast inference of mixture-of-experts language models with offloading},
  author={Eliseev, Artyom and Mazur, Denis},
  journal={arXiv preprint arXiv:2312.17238},
  year={2023}
}

@article{xue2024moe,
  title={Moe-infinity: Offloading-efficient moe model serving},
  author={Xue, Leyang and Fu, Yao and Lu, Zhan and Mai, Luo and Marina, Mahesh},
  journal={arXiv preprint arXiv:2401.14361},
  year={2024}
}

@article{song2024promoe,
  title={Promoe: Fast moe-based llm serving using proactive caching},
  author={Song, Xiaoniu and Zhong, Zihang and Chen, Rong and Chen, Haibo},
  journal={arXiv preprint arXiv:2410.22134},
  year={2024}
}

@inproceedings{yu2026taming,
  title={Taming latency-memory trade-off in MoE-based LLM serving via fine-grained expert offloading},
  author={Yu, Hanfei and Cui, Xingqi and Zhang, Hong and Wang, Hao and Wang, Hao},
  booktitle={Proceedings of the 21st European Conference on Computer Systems},
  pages={176--191},
  year={2026}
}

@inproceedings{cao2025moe,
  title={Moe-lightning: High-throughput moe inference on memory-constrained gpus},
  author={Cao, Shiyi and Liu, Shu and Griggs, Tyler and Schafhalter, Peter and Liu, Xiaoxuan and Sheng, Ying and Gonzalez, Joseph E and Zaharia, Matei and Stoica, Ion},
  booktitle={Proceedings of the 30th ACM International Conference on Architectural Support for Programming Languages and Operating Systems, Volume 1},
  pages={715--730},
  year={2025}
}

@inproceedings{kamahori2025fiddler,
  title={Fiddler: Cpu-gpu orchestration for fast inference of mixture-of-experts models},
  author={Kamahori, Keisuke and Tang, Tian and Gu, Yile and Zhu, Kan and Kasikci, Baris},
  booktitle={International Conference on Learning Representations},
  volume={2025},
  pages={56099--56115},
  year={2025}
}

@inproceedings{zhong2025hybrimoe,
  title={Hybrimoe: Hybrid cpu-gpu scheduling and cache management for efficient moe inference},
  author={Zhong, Shuzhang and Sun, Yanfan and Liang, Ling and Wang, Runsheng and Huang, Ru and Li, Meng},
  booktitle={2025 62nd ACM/IEEE Design Automation Conference (DAC)},
  pages={1--7},
  year={2025},
  organization={IEEE}
}

@inproceedings{zhang2025daop,
  title={Daop: Data-aware offloading and predictive pre-calculation for efficient moe inference},
  author={Zhang, Yujie and Aggarwal, Shivam and Mitra, Tulika},
  booktitle={2025 Design, Automation \& Test in Europe Conference (DATE)},
  pages={1--7},
  year={2025},
  organization={IEEE}
}

@inproceedings{chen2025ktransformers,
  title={Ktransformers: Unleashing the full potential of cpu/gpu hybrid inference for moe models},
  author={Chen, Hongtao and Xie, Weiyu and Zhang, Boxin and Tang, Jingqi and Wang, Jiahao and Dong, Jianwei and Chen, Shaoyuan and Yuan, Ziwei and Lin, Chen and Qiu, Chengyu and others},
  booktitle={Proceedings of the ACM SIGOPS 31st Symposium on Operating Systems Principles},
  pages={1014--1029},
  year={2025}
}

@article{wang2024model,
  title={Model compression and efficient inference for large language models: A survey},
  author={Wang, Wenxiao and Chen, Wei and Luo, Yicong and Long, Yongliu and Lin, Zhengkai and Zhang, Liye and Lin, Binbin and Cai, Deng and He, Xiaofei},
  journal={arXiv preprint arXiv:2402.09748},
  year={2024}
}

@article{liu2026survey,
  title={A survey on inference optimization techniques for mixture of experts models},
  author={Liu, Jiacheng and Tang, Peng and Wang, Wenfeng and Ren, Yuhang and Hou, Xiaofeng and Heng, Pheng Ann and Guo, Minyi and Li, Chao},
  journal={ACM Computing Surveys},
  volume={58},
  number={10},
  pages={1--37},
  year={2026},
  publisher={ACM New York, NY}
}

@article{fedus2022switch,
  title={Switch transformers: Scaling to trillion parameter models with simple and efficient sparsity},
  author={Fedus, William and Zoph, Barret and Shazeer, Noam},
  journal={Journal of Machine Learning Research},
  volume={23},
  number={120},
  pages={1--39},
  year={2022}
}

@article{jiang2024mixtral,
  title={Mixtral of experts},
  author={Jiang, Albert Q and Sablayrolles, Alexandre and Roux, Antoine and Mensch, Arthur and Savary, Blanche and Bamford, Chris and Chaplot, Devendra Singh and Casas, Diego de las and Hanna, Emma Bou and Bressand, Florian and others},
  journal={arXiv preprint arXiv:2401.04088},
  year={2024}
}

@inproceedings{kwon2023efficient,
  title={Efficient memory management for large language model serving with pagedattention},
  author={Kwon, Woosuk and Li, Zhuohan and Zhuang, Siyuan and Sheng, Ying and Zheng, Lianmin and Yu, Cody Hao and Gonzalez, Joseph and Zhang, Hao and Stoica, Ion},
  booktitle={Proceedings of the 29th symposium on operating systems principles},
  pages={611--626},
  year={2023}
}

@misc{touvron2023llamaopenefficientfoundation,
      title={LLaMA: Open and Efficient Foundation Language Models}, 
      author={Hugo Touvron and Thibaut Lavril and Gautier Izacard and Xavier Martinet and Marie-Anne Lachaux and Timothée Lacroix and Baptiste Rozière and Naman Goyal and Eric Hambro and Faisal Azhar and Aurelien Rodriguez and Armand Joulin and Edouard Grave and Guillaume Lample},
      year={2023},
      eprint={2302.13971},
      archivePrefix={arXiv},
      primaryClass={cs.CL},
      url={https://arxiv.org/abs/2302.13971}, 
}

@article{merity2016pointer,
  title={Pointer sentinel mixture models},
  author={Merity, Stephen and Xiong, Caiming and Bradbury, James and Socher, Richard},
  journal={arXiv preprint arXiv:1609.07843},
  year={2016}
}

@article{marcus1993building,
  title={Building a large annotated corpus of English: The Penn Treebank},
  author={Marcus, Mitch and Santorini, Beatrice and Marcinkiewicz, Mary Ann},
  journal={Computational linguistics},
  volume={19},
  number={2},
  pages={313--330},
  year={1993}
}

@article{raffel2020exploring,
  title={Exploring the limits of transfer learning with a unified text-to-text transformer},
  author={Raffel, Colin and Shazeer, Noam and Roberts, Adam and Lee, Katherine and Narang, Sharan and Matena, Michael and Zhou, Yanqi and Li, Wei and Liu, Peter J},
  journal={Journal of machine learning research},
  volume={21},
  number={140},
  pages={1--67},
  year={2020}
}

@inproceedings{mihaylov2018can,
  title={Can a suit of armor conduct electricity? a new dataset for open book question answering},
  author={Mihaylov, Todor and Clark, Peter and Khot, Tushar and Sabharwal, Ashish},
  booktitle={Proceedings of the 2018 conference on empirical methods in natural language processing},
  pages={2381--2391},
  year={2018}
}

@article{clark2018think,
  title={Think you have solved question answering? try arc, the ai2 reasoning challenge},
  author={Clark, Peter and Cowhey, Isaac and Etzioni, Oren and Khot, Tushar and Sabharwal, Ashish and Schoenick, Carissa and Tafjord, Oyvind},
  journal={arXiv preprint arXiv:1803.05457},
  year={2018}
}

@article{sakaguchi2021winogrande,
  title={Winogrande: An adversarial winograd schema challenge at scale},
  author={Sakaguchi, Keisuke and Bras, Ronan Le and Bhagavatula, Chandra and Choi, Yejin},
  journal={Communications of the ACM},
  volume={64},
  number={9},
  pages={99--106},
  year={2021},
  publisher={ACM New York, NY, USA}
}

@inproceedings{zellers2019hellaswag,
  title={Hellaswag: Can a machine really finish your sentence?},
  author={Zellers, Rowan and Holtzman, Ari and Bisk, Yonatan and Farhadi, Ali and Choi, Yejin},
  booktitle={Proceedings of the 57th annual meeting of the association for computational linguistics},
  pages={4791--4800},
  year={2019}
}

@inproceedings{bisk2020piqa,
  title={Piqa: Reasoning about physical commonsense in natural language},
  author={Bisk, Yonatan and Zellers, Rowan and Gao, Jianfeng and Choi, Yejin and others},
  booktitle={Proceedings of the AAAI conference on artificial intelligence},
  volume={34},
  number={05},
  pages={7432--7439},
  year={2020}
}

@article{lai2026rs,
  title={RS-MoE: Coupled expert compression via activation-peak guided collaborative decomposition},
  author={Lai, Jiajun and Li, Shijie and Li, Haoqin and Xue, Ying and Jiang, Huaiguang},
  journal={Knowledge-Based Systems},
  pages={116244},
  year={2026},
  publisher={Elsevier}
}

@inproceedings{li2026sub,
  title={Sub-moe: Efficient mixture-of-expert llms compression via subspace expert merging},
  author={Li, Lujun and Zhu, Qiyuan and Wang, Jiacheng and Qin, Xiaoyu and Li, Wei and Gu, Hao and Han, Sirui and Guo, Yike},
  booktitle={Proceedings of the AAAI Conference on Artificial Intelligence},
  volume={40},
  number={27},
  pages={22994--23002},
  year={2026}
}

@inproceedings{xu2026camera,
  title={CAMERA: Multi-matrix joint compression for MoE models via micro-expert redundancy analysis},
  author={Xu, Yuzhuang and Han, Xu and Zhang, Yuanchi and Wang, Yixuan and Liu, Yijun and Ji, Shiyu and Zhu, Qingfu and Che, Wanxiang},
  booktitle={Proceedings of the AAAI Conference on Artificial Intelligence},
  volume={40},
  number={32},
  pages={27395--27404},
  year={2026}
}

@inproceedings{wang2025svd2,
  title={Svd-llm v2: Optimizing singular value truncation for large language model compression},
  author={Wang, Xin and Alam, Samiul and Wan, Zhongwei and Shen, Hui and Zhang, Mi},
  booktitle={Proceedings of the 2025 Conference of the Nations of the Americas Chapter of the Association for Computational Linguistics: Human Language Technologies (Volume 1: Long Papers)},
  pages={4287--4296},
  year={2025}
}

@inproceedings{wang2025dobi,
  title={Dobi-svd: Differentiable SVD for LLM compression and some new perspectives},
  author={Wang, Qinsi and Ke, Jinghan and Tomizuka, Masayoshi and Keutzer, Kurt and Xu, Chenfeng},
  booktitle={International Conference on Learning Representations},
  volume={2025},
  pages={12561--12590},
  year={2025}
}

@article{bai2025erc,
  title={ERC-SVD: Error-Controlled SVD for Large Language Model Compression},
  author={Bai, Haolei and Jian, Siyong and Liang, Tuo and Yin, Yu and Wang, Huan},
  journal={arXiv preprint arXiv:2505.20112},
  year={2025}
}

@inproceedings{amini2019mathqa,
  title={Mathqa: Towards interpretable math word problem solving with operation-based formalisms},
  author={Amini, Aida and Gabriel, Saadia and Lin, Shanchuan and Koncel-Kedziorski, Rik and Choi, Yejin and Hajishirzi, Hannaneh},
  booktitle={Proceedings of the 2019 Conference of the North American Chapter of the Association for Computational Linguistics: Human Language Technologies, Volume 1 (Long and Short Papers)},
  pages={2357--2367},
  year={2019}
}

@inproceedings{yang2024moe,
  title={Moe-i2: Compressing mixture of experts models through inter-expert pruning and intra-expert low-rank decomposition},
  author={Yang, Cheng and Sui, Yang and Xiao, Jinqi and Huang, Lingyi and Gong, Yu and Duan, Yuanlin and Jia, Wenqi and Yin, Miao and Cheng, Yu and Yuan, Bo},
  booktitle={Findings of the Association for Computational Linguistics: EMNLP 2024},
  pages={10456--10466},
  year={2024}
}

@article{huang2025milo,
  title={Milo: Efficient quantized moe inference with mixture of low-rank compensators},
  author={Huang, Beichen and Yuan, Yueming and Shao, Zelei and Zhang, Minjia},
  journal={Proceedings of Machine Learning and Systems},
  volume={7},
  year={2025}
}

@inproceedings{zhong2024adapmoe,
  title={Adapmoe: Adaptive sensitivity-based expert gating and management for efficient moe inference},
  author={Zhong, Shuzhang and Liang, Ling and Wang, Yuan and Wang, Runsheng and Huang, Ru and Li, Meng},
  booktitle={Proceedings of the 43rd IEEE/ACM International Conference on Computer-Aided Design},
  pages={1--9},
  year={2024}
}

@article{lyu2026spice,
  title={SPICE: Speculative Prefetching with Low-Rank Expert Surrogates and Heterogeneous Orchestration for MoE Inference Acceleration},
  author={Lyu, Yongxiang and Li, Ning and Jia, Bonian},
  journal={arXiv preprint arXiv:2608.21240},
  year={2026}
}

\appendix

\end{document}